\documentclass[floats,floatfix,showpacs,amssymb,prd,twocolumn,superscriptaddress,nofootinbib,nolongbibliography,reprint]{revtex4-1}
	
\usepackage{amssymb,amsmath,verbatim,mathtools,needspace,enumitem,etoolbox,graphicx,physics,microtype,afterpage}
\usepackage{gensymb}

\usepackage[dvipsnames, usenames]{xcolor}

\definecolor{linkcolor}{rgb}{0.0,0.3,0.5}
\usepackage[unicode, colorlinks=true, linkcolor=linkcolor, citecolor=linkcolor, filecolor=linkcolor,urlcolor=linkcolor, pdfusetitle]{hyperref}
\usepackage[all]{hypcap}
\usepackage[T1]{fontenc}
\usepackage[utf8]{inputenc}
\usepackage{tabularx}

\newcommand{\ssim}{\mathchar"5218\relax\,}

\renewcommand{\vec}[1]{\mathbf{#1}}

\def\apj{\rm{ApJ}}
\def\apjl{\rm{ApJ}}

\def\aap{\rm{A\&A}}

\def\mnras{\rm{MNRAS}}
\def\prd{\rm{PRD}}
\def\prl{\rm{PRL}}
\def\prx{\rm{PRX}}

\def\cqg{\rm{CQG}}
\def\lrr{\rm{LRR}}

\newcommand{\bham}{\affiliation{School of Physics and Astronomy \& Institute for Gravitational Wave Astronomy, \\ University of Birmingham, Birmingham, B15 2TT, United Kingdom}}

\makeatletter
\newcommand*{\balancecolsandclearpage}{%
  \close@column@grid
  \cleardoublepage
  \twocolumngrid
}
\makeatother

\begin{document}

\title{Endpoint of the up-down instability in precessing binary black holes}

\author{Matthew Mould}
\email{mmould@star.sr.bham.ac.uk}
 \bham

\author{Davide Gerosa}
\email{d.gerosa@bham.ac.uk}
 \bham


\date{\today}

\begin{abstract}
Binary black holes in which both spins are aligned with the
binary's
orbital angular momentum do not precess.
However, the up-down configuration, in which the spin of the heavier (lighter) black hole is aligned (anti-aligned) with the orbital angular momentum, is unstable to spin precession at small orbital separations~[D. Gerosa {\it et al.}, \href{https://doi.org/10.1103/PhysRevLett.115.141102}{Phys. Rev. Lett. 115, 141102 (2015)}].
We first cast the spin precession problem in terms of a simple harmonic oscillator and provide a cleaner derivation of the instability onset.
Surprisingly, we find that following the instability, up-down binaries do not disperse in the available parameter space but evolve toward precise endpoints. 
We then present an analytic scheme to locate these final configurations and confirm them with numerical integrations.
Namely, unstable up-down binaries approach mergers with the two spins coaligned with each other and equally misaligned with the orbital angular momentum. 
Merging up-down binaries relevant to LIGO/Virgo and LISA may be detected in these endpoint configurations if the instability onset occurs prior to the sensitivity threshold of the detector.
As a by-product, we obtain new generic results on binary black hole spin-orbit resonances at 2nd~post-Newtonian order.
We finally apply these findings to a simple astrophysical population of binary black holes where a formation mechanism aligns the spins without preference for co- or counteralignment, as might be the case for stellar-mass black holes embedded in the accretion disk of a supermassive black hole.
\end{abstract}

\maketitle

\section{Introduction} \label{Introduction}

Stellar-mass black hole (BH) binaries are now regularly detected
by the gravitational-wave (GW) detectors LIGO
and Virgo~\cite{2019PhRvX...9c1040A}.
LISA will soon observe supermassive BH binaries which populate the low-frequency GW sky~\cite{2017arXiv170200786A}.
These detections provide the opportunity to study BHs as never before, allowing for the confrontation of theory with observation.
The evolution of binary BHs generalizes the Newtonian two-body problem to Einstein's theory of general relativity.
Though no exact solution is known, several approximate methods have been developed to tackle this problem, including the post-Newtonian (PN)~\cite{2014LRR....17....2B},
effective-one-body~\cite{1999PhRvD..59h4006B}, and gravitational self-force~\cite{ 2011LRR....14....7P} formalisms, as well as numerical relativity~\cite{2019RPPh...82a6902D}.

The simplest system one can address is that of two nonspinning BHs.
Beyond this is the case in which the holes have spins aligned with the orbital angular momentum of the binary.
These configurations are unique among spinning BH binaries in that such a system does not precess: the orbital plane maintains a fixed orientation and their gravitational emission is comparatively easy to model. 
For generic  sources in which the BH spins are misaligned, the orbital angular momentum and both BH spins all precess about the total angular momentum.
The resulting relativistic spin-orbit and spin-spin couplings~\cite{1995PhRvD..52..821K} give rise to a very rich precessional dynamics, leading to modulations in the emitted gravitational waveform~\cite{1985PhRvD..31.1815T, 1994PhRvD..49.6274A}.
Accurate modeling of spin precession is crucial to interpret current and future GW events~\cite{2017PhRvD..95j4038C,2017PhRvD..95b4010B,2019PhRvD.100b4059K,2019PhRvR...1c3015V}

Spins are clean astrophysical observables. For stellar-mass BHs observed by LIGO/Virgo, they are a powerful tools to discriminate between isolated and dynamically assembled binaries~\cite{2014PhRvL.112y1101V,2014PhRvD..89l4025G,2016ApJ...832L...2R,2017MNRAS.471.2801S,2019ApJ...886...25B}. BH spins encode information on some essential physics of massive stars including, but not limited to, core-envelope interactions, tides, mass transfer, supernova kicks, magnetic torquing, and internal gravity waves~\citep{
2013PhRvD..87j4028G,
2013ApJ...764..166D,
2015ApJ...810..101F,
2020A&A...636A.104B,
2018ApJ...862L...3S, 
2018A&A...616A..28Q,
2018PhRvD..98h4036G,
2018MNRAS.473.4174Z,
2020A&A...635A..97B,
2019ApJ...881L...1F,
2020ApJ...894..129S}. For binaries embedded in gaseous environments such as the disks of active galactic nuclei (AGN)~\cite{2017MNRAS.464..946S, 2017ApJ...835..165B}, spin misalignments might allow us to constrain the occurrence of relativistic viscous interactions \cite{1975ApJ...195L..65B}. This is also the case for supermassive BH binaries that populate the LISA band, where prominent phases of disk accretion might crucially impact the spin orientations at merger \cite{2008ApJ...684..822B,2009MNRAS.399.2249P,2014ApJ...794..104S,2015MNRAS.451.3941G,2016MNRAS.456..961B}.

\begin{figure*}[t]
\centering
\includegraphics[width=1.0\textwidth]{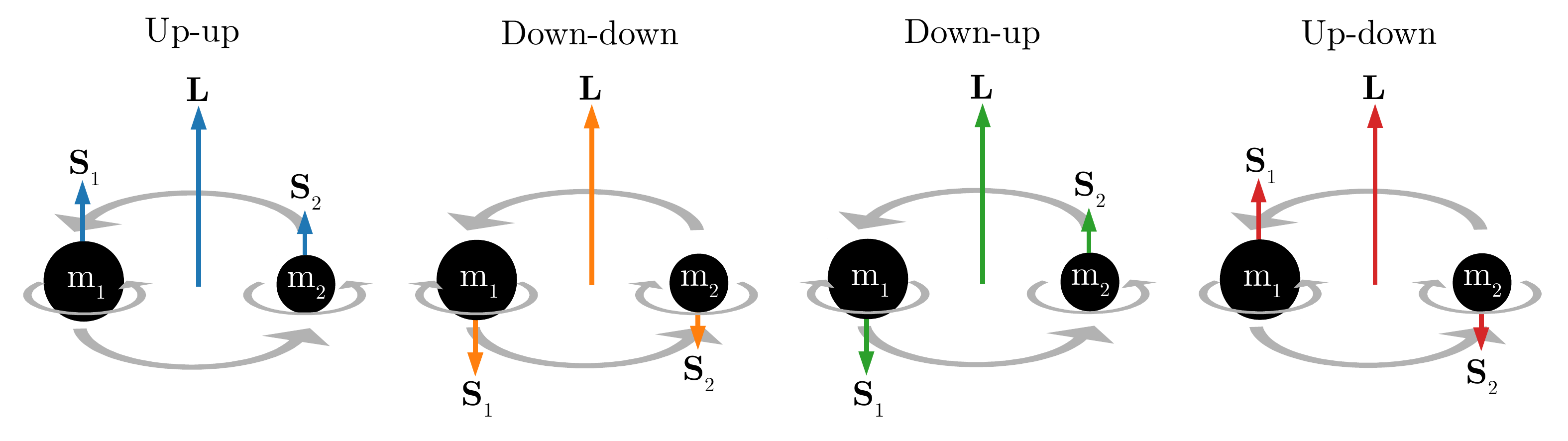}
\caption{The four binary BH configurations with aligned spins. The BH with higher (lower) mass is indexed by the number 1 (2). We refer to the orientation of a BH whose spin vector $\vec{S}_i$ is parallel (antiparallel) to the orbital angular momentum vector $\vec{L}$ as ``up'' (``down''). The four distinct binary configurations are then labeled with the orientation of the primary (secondary) BH appearing before (after) the hyphen.}
\label{fig aligned}
\end{figure*}

There are four distinct configurations in which the BH spins are aligned to the orbital angular momentum (see Fig.~\ref{fig aligned}).
We dub each of these cases ``up-up'', ``down-down'', ``down-up'' and ``up-down'', where ``up'' (``down'') refers to co- (counter-) alignment with the orbital angular momentum and the label before (after) the hyphen refers to the spin alignment of the primary (secondary) BH. It is straightforward to show that all four of these configurations are equilibrium, nonprecessing solutions of the relativistic spin-precession equations~\cite{1995PhRvD..52..821K}: a BH binary initialized in \emph{exactly} one of these configurations remains so over its inspiral.
Here, we tackle their stability: if an arbitrarily small misalignment is present, how do such configurations behave?

Employing the parametrization of generic spin precession in terms of an effective potential at 2PN order~\cite{2015PhRvL.114h1103K, 2015PhRvD..92f4016G}, \citeauthor{2015PhRvL.115n1102G} \cite{2015PhRvL.115n1102G} investigated the robustness of aligned spin binary BH configurations (see also Ref.~\cite{2016PhRvD..93l4074L} for a subsequent study).
They found that the up-up, down-down and down-up configurations are stable, remaining approximately aligned under a small perturbation of the spin directions.
This is not the case for up-down binaries, i.e. those where the heavier BH is aligned with the orbital angular momentum while the lighter BH is antialigned. They report the presence of a critical orbital separation
\begin{align} \label{rudintro}
r_{\mathrm{ud}+} = \frac{\left( \sqrt{\chi_1} + \sqrt{q\chi_2} \right)^4}{(1-q)^2} M
\end{align}
which defines the onset of the instability (here $q < 1$ is the binary mass ratio, $M$ is the total mass, $\chi_1$ and $\chi_2$ are the Kerr parameters of the more and less massive BH, respectively, and we use geometrical units $G=c=1$).
An up-down binary that is formed at large orbital separations $r>r_{\mathrm{ud}+}$ will at first inspiral much as the other stable aligned binaries do, with the spins remaining arbitrarily close to the aligned configuration.
However, upon reaching the instability onset at $r=r_{\mathrm{ud}+}$, the binary becomes unstable to spin precession, leading to large misalignments of the spins.

\begin{figure}
\centering
\includegraphics[width=0.9\columnwidth]{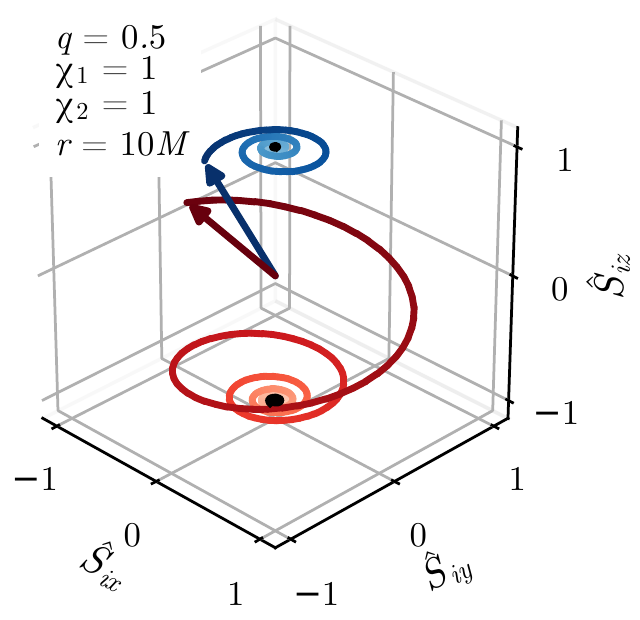}
\caption{Numerical evolution of the normalized spins $\hat{\vec{S}}_i = \vec{S}_i/S_i$ of a BH binary with mass ratio $q = 0.5$ and dimensionless spins $\chi_1 = \chi_2 = 1$. The blue (red) curve traces the path of the spin vector $\vec{S}_1$ ($\vec{S}_2$) of the heavier (lighter) BH over the evolution. The integration is performed from a binary separation $r = 1000M$ to $10M$; the colors of the curves darken with decreasing separation.
The binary is initialized with misalignments of $1\degree$ in the BH spins from the up-down configuration. The vertical $z$-axis is initially aligned to the total angular momentum, the $x$-axis is constructed such that the initial orbital angular momentum lies in the $x$-$z$ plane, and the $y$-axis completes the orthogonal frame.
 The black dots show the location of the spins for $r > r_{\mathrm{ud}+} \simeq 34 M$, before the onset of instability. The arrows show the orientation of the spins at the final separation $r=10M$. The binary is approaching the endpoint listed in Eq.~(\ref{endpointintro}). An animated version of this figure is available at \href{https://davidegerosa.com/spinprecession/}{www.davidegerosa.com/spinprecession}.}
\label{fig inertial}
\end{figure}

Figure~\ref{fig inertial} shows the evolution of the spins for a binary BH in the up-down configuration.
The binary is evolved from an orbital separation of $r=1000M>r_{\rm ud+}$ to $r=10M$.
At the initial separation, the spin directions are perturbed such that there is a misalignment of $1\degree$ in the spins from the exact up-down configuration.
The response to this perturbation is initially tight polar oscillations (black dots in Fig.~\ref{fig inertial}) of the BH spins around the aligned configuration.
After the onset of instability, precession induces large spin misalignments (colored tracks in Fig.~\ref{fig inertial}).

A key question so far unanswered is the following: after becoming unstable, to what configuration do up-down binaries evolve? In other words: \emph{what is the endpoint of the up-down instability?}

In this paper, we present a detailed study on the onset and evolution of unstable up-down binary BHs.
In Sec.~\ref{Instability threshold} we provide a novel derivation of the stability onset directly from the orbit-averaged 2PN spin precession equations.
We test the robustness of the result with numerical PN evolutions of BH binaries and find that unstable binaries tend to cluster in specific locations of the parameter space by the end of their evolutions.
In Sec.~\ref{Resonant configurations} we explore this observation analytically.
Previous investigations \cite{2015PhRvL.115n1102G} highlighted connections between the up-down instability and the so-called spin-orbit resonances~\cite{2004PhRvD..70l4020S} -- peculiar BH binary configurations where the two spins and the angular momentum remain coplanar. We present a new semianalytic scheme to locate the resonances and confirm that the evolution of the up-down instability is inherently connected to the nature of these configurations.

We obtain a surprisingly simple result (Sec.~\ref{Up-down endpoint}): after undergoing the instability, up-down binaries tend to the very special configuration where the two BH spins $\vec{S}_1$ and $\vec{S}_2$ are coaligned with each other and equally misaligned with the orbital angular momentum $\vec{L}$. More specifically, the endpoint of the up-down instability is a precessing configuration with (using hats to denote unit vectors)
\begin{align}
\label{endpointintro}
\hat{\vec S}_1= \hat{\vec S}_2 \quad\mathrm{and} \quad 
\hat{\vec S}_1 \cdot \hat{\vec L} = \hat{\vec S}_2 \cdot \hat{\vec L} = \frac{\chi_1-q\chi_2}{\chi_1+q\chi_2} \, .
\end{align}
From the distribution of endpoints of populations of up-down binaries, we characterize the typical conditions required for such binaries to become unstable before the end of their evolutions and the typical growth time of the precessional instability. We then explore the astrophysical relevance of our finding for a population of stellar-mass BH binaries formed in AGN disks, and finally draw our concluding remarks (Sec.~\ref{Conclusions}).

\section{Instability threshold} \label{Instability threshold}

\subsection{2PN binary black hole spin precession}

We denote vectors in bold, e.g. $\vec{v}$, magnitudes with $v=|\vec{v}|$, and unit vectors with $\hat{\vec{v}}$.
Throughout the paper we use geometrical units $G = c = 1$. Let us consider binary BHs with component masses $m_1$ and $m_2$, total mass $M = m_1 + m_2$, mass ratio $q = m_2/m_1 \leq 1$ and symmetric mass ratio $\eta = q/(1+q)^2$.
We denote the binary separation with $r$ and the Newtonian angular momentum with $L=\eta(M^3r)^{1/2}$.
The spins of the two BHs are denoted by $\vec{S}_i = m_i^2\chi_i\hat{\vec{S}}_i$ ($i=1,2$), where $0 \leq \chi_i \leq 1$ are the dimensionless Kerr parameters. The total spin is $\vec{S} = \vec{S}_1 + \vec{S}_2$ and the total angular momentum is $\vec{J} = \vec{L} + \vec{S}$. We consider orbital separations $r \geq 10M$, which is taken as the breakdown of the PN approximation \cite{2006PhRvD..74j4005B, 2009PhRvD..79h4010C, 2009PhRvD..80h4043B}.

There are three timescales on which generically precessing binary BHs evolve:
\begin{itemize}
\item the orbital timescale, given by the Keplarian expression $t_\mathrm{orb}/M \simeq (r/M)^{3/2}$, on which the BHs orbit each other,
\item the precession timescale, $t_\mathrm{orb}/M \simeq (r/M)^{5/2}$, on which $\vec{S}_1$, $\vec{S}_2$, and $\vec{L}$ change direction~\cite{1994PhRvD..49.6274A}, and
\item the radiation-reaction timescale, $t_\mathrm{orb}/M \simeq (r/M)^{4}$, on which the binary separation shrinks due to GW emission~\cite{1964PhRv..136.1224P}.
\end{itemize}
In the post-Newtonian (PN) regime $r\gg M$ these timescales are separated, so that 
\begin{equation}
t_\mathrm{orb} \ll t_\mathrm{pre} \ll t_\mathrm{RR}\,.
\end{equation}
The BHs orbit each other many times before completing one precession cycle, and complete many precession cycles before the binary separation decreases. This hierarchy of timescales allows each part of the binary dynamics -- the orbital, precessional, and radiation-reaction motion -- to be addressed independently.
The inequality $t_\mathrm{orb} \ll t_\mathrm{pre}$ has been used to study precession in binary BHs by averaging the motion over the orbital period (e.g., \cite{2004PhRvD..70l4020S,2008PhRvD..78d4021R}). Further, the inequality $t_\mathrm{pre} \ll t_\mathrm{RR}$ has been used to separate the precessional motion from the GW-driven inspiral~\cite{2015PhRvL.114h1103K, 2015PhRvD..92f4016G,2017CQGra..34f4004G,2017PhRvD..96b4007Z,2019CQGra..36j5003G}.

The 2PN orbit-averaged equations describing the evolutions of the BH spins and the orbital angular momentum read~\cite{2008PhRvD..78d4021R}
\begin{subequations}
\label{precession equations}
\begin{align}
\label{dS1/dt}
\frac{d\vec{S}_1}{dt} = {}& \frac{1}{2r^3} \bigg\{ \left[ 4 + 3q - \frac {3M^2q\xi} {(1+q)L} \right] \vec{L} + \vec{S}_2 \bigg\} \times \vec{S}_1
\, , \\
\label{dS2/dt}
\frac{d\vec{S}_2}{dt} = {}& \frac{1}{2r^3} \bigg\{ \left[ 4+\frac{3}{q} - \frac {3M^2\xi} {(1+q)L} \right] \vec{L} + \vec{S}_1 \bigg\} \times \vec{S}_2
\, , \\
\label{dL/dt} \nonumber
\frac{d\vec{L}}{dt} = {}& \frac{1}{2r^3} \bigg\{ \left[ 4 + 3q - \frac {3M^2q\xi} {(1+q)L} \right] \vec{S}_1
\\
&\! + \left[ 4 + \frac{3}{q} - \frac {3M^2\xi} {(1+q)L} \right] \vec{S}_2 \bigg\} \times \vec{L} + \frac{dL}{dt} \hat{\vec{L}}
\, ,
\end{align}
\end{subequations}
where $\xi$ is the projected effective spin (often referred to as $\chi_\mathrm{eff}$ \cite{2019PhRvX...9c1040A,2019ApJ...882L..24A}),
\begin{align} \label{xi}
\xi = \frac{1}{M^2} \left[ (1+q) \vec{S}_1 + \left( 1+\frac{1}{q} \right) \vec{S}_2 \right] \cdot \hat{\vec{L}}
\, . 
\end{align}
On the precessional timescale, $dL/dt\simeq 0$ and the evolutionary equations describe precessional motions of the three vectors $\vec{L}$, $\vec{S}_1$,  and $\vec{S}_2$ about $\vec{J}$. The evolution on the longer radiation-reaction timescale is supplemented by a PN equation for $dL/dt$. In this paper we include (non) spinning terms up to 3.5PN (2PN); cf. e.g. Eq.~(27) in Ref.~\cite{2016PhRvD..93l4066G}.

The effective spin $\xi$ is a constant of motion of the orbit-averaged problem at 2PN in spin precession and 2.5PN in radiation reaction~\cite{2008PhRvD..78d4021R}. The magnitudes $S_1$ and $S_2$ of the BHs spins are also constant. On the short precessional timescale, the separation $r$ and total angular momentum 
\begin{equation}
J = |\vec{L} + \vec{S}_1 + \vec{S}_2|
\end{equation} are conserved. The entire precessional dynamics can be parametrized with a single variable, the total spin magnitude~\cite{2015PhRvL.114h1103K,2015PhRvD..92f4016G}
\begin{equation}
S = |\vec{S}_1 + \vec{S}_2|
\, .
\end{equation}
Excluding the case of transitional precession where $J\ssim 0$~\cite{1994PhRvD..49.6274A}, the direction $\hat{\vec{J}}$ is conserved to very high accuracy also on the longer radiation-reaction timescale~\cite{2017PhRvD..96b4007Z}.

\begin{figure}[t]
\centering
\includegraphics[width=0.8\columnwidth]{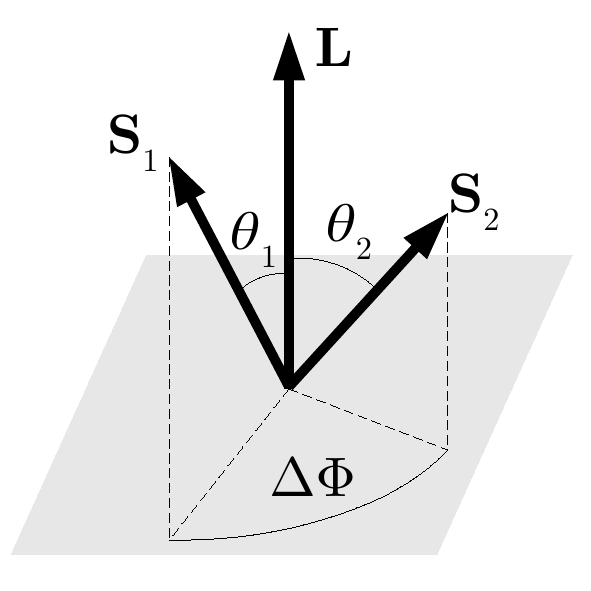}
\caption{The noninertial frame aligned with the orbital angular momentum $\vec{L}$. The angle between each spin vector $\vec{S}_i$ and $\vec{L}$ is denoted by $\theta_i$ ($i=1,2$), while
the angle between the projections of the two spins onto the orbital plane is denoted by $\Delta\Phi$.}
\label{fig vectors}
\end{figure}

In a noninertial frame coprecessing with $\hat{\vec{L}}$, we define the relative orientations of the spin directions by the angles $\theta_i$ between $\hat{\vec{S}}_i$ and $\hat{\vec{L}}$ and the angle $\Delta\Phi$ between the projections of the spins onto the orbital plane (see Fig.~\ref{fig vectors} for a schematic representation):
\begin{subequations}
\label{cos}
\begin{align}
\label{cos theta1}
\cos{\theta_1} &= \hat{\vec{S}}_1 \cdot\hat{\vec{L}} 
\, ,\\
\label{cos theta2}
\qquad \cos{\theta_2} &= \hat{\vec{S}}_2 \cdot\hat{\vec{L}}
\, , \\
\label{cos DeltaPhi}
\cos{\Delta\Phi} &=\frac{\hat{\vec{S}}_1 \cross \hat{\vec{L}}}{|\hat{\vec{S}}_1 \cross \hat{\vec{L}}|} \cdot \frac{\hat{\vec{S}}_2 \cross \hat{\vec{L}}}{|\hat{\vec{S}}_2 \cross \hat{\vec{L}}|}
\, .
\end{align}
\end{subequations}
For given values of $q$, $\chi_1$, and $\chi_2$, 
the mutual orientations of the three vectors $\vec{L}$, $\vec{S}_1$, and $\vec{S}_2$ can be parametrized equivalently  in terms of either ($\xi,J,S$) or $(\theta_1,\theta_2,\Delta\Phi)$. The conversion between the two sets of variables is given explicitly in Eqs.~(8-9) of Ref.~\cite{2016PhRvD..93l4066G}.

\subsection{Binary black hole spins as harmonic oscillators} \label{Binary BH spins as harmonic oscillators}

One can immediately prove that binaries with aligned spins $\hat{\vec{L}}=\hat{\vec{S}}_1=\hat{\vec{S}}_2$ are equilibrium solutions of Eqs.~(\ref{dS1/dt}-\ref{dL/dt}). The stability of the solutions is determined by their response to small perturbations. The investigations of Ref.~\cite{2015PhRvL.115n1102G} indicate that the up-down instability develops on the short precessional timescale $t_{\rm pre}$. In this regime, all variables can be kept constant but $S$.

The evolution of $S$ is determined directly by Eqs.~(\ref{dS1/dt}-\ref{dS2/dt}):
\begin{align} \label{dS/dt}
\frac{dS^2}{dt} = A \sqrt{a_6S^6 + a_4S^4 + a_2S^2 + a_0}
\, ,
\end{align}
where
\begin{subequations}
\begin{align}
A ={}&\! -\frac{3(1-q^2)}{q} S_1S_2 \left( \frac{M^3\eta^2}{L^2} \right)^3 \left( L-M^2\eta\xi \right) \, , \\
\label{coeffa6}
a_6 ={}&\! -\frac{q}{4(1-q)^2 S_1^2S_2^2 L^2} \, , \\
\label{coeffa4}
a_4 ={}&\! -a_6q^{-1} \big[ \big(1+q^2\big)L^2 - 2qJ^2 + 2M^2q\xi L \notag \\
&\! -(1-q)\big(qS_1^2-S_2^2\big)\big] \, , \\
\label{coeffa2}
a_2 ={}& a_6q^{-1} \big\{ q(1+q)^2J^4 \notag \\
&\! - 2(1+q)^2J^2 \big[ qL^2 + M^2q\xi L - (1-q)\big(qS_1^2-S_2^2\big) \big] \notag \\
&\! + (1+q)^2L^2 \big[ qL^2 - 2(1-q)\big(S_1^2-qS_2^2\big) \big] \notag \\
&\! + 2(1+q)M^2q\xi L \big[ (1+q)L^2 - (1-q)\big(S_1^2-S_2^2\big) \big] \notag \\
&\! + 4M^4q^2\xi^2 L^2 \big\} \, , \\
\label{coeffa0}
a_0 ={}& a_6q^{-1} \big\{ (1+q)J^4\big(qS_1^2-S_2^2\big) \notag \\
&\! -2J^2 \big[ (1+q)\big(qS_1^2-S_2^2\big)L^2 + \big(S_1^2-S_2^2)M^2q\xi L \big] \notag \\
&\! -L^2 \big[ \big(1-q^2\big)\big(S_1^2-S_2^2\big)^2 - (1+q)\big(qS_1^2-S_2^2\big)L^2 \notag \\
&\! -2M^2q\xi L\big(S_1^2-S_2^2\big) \big] \big\} \, .
\end{align}
\end{subequations}
The conservation of $\xi$, $J$, $L$, $S_1$, and $S_2$ over $t_\mathrm{pre}$ implies that, after taking a second time derivative, only the derivatives of $S^2$ survive and Eq.~(\ref{dS/dt}) becomes
\begin{align}
\label{anharmonic}
\frac{d^2S^2}{dt^2}  = \frac{A^2}{2} \left(3 a_6 S^4 +2 a_4 S^2 + a_2 \right)
\, .
\end{align}
By rearranging the right-hand side of Eq.~(\ref{anharmonic}) we find that the time evolution for a perturbation $S^2 - S_*^2$ to some solution $S_*$ of Eq.~(\ref{dS/dt}) is determined by
\begin{align} \label{perturbation}
\frac{d^2 (S^2 - S_*^2)}{dt^2} ={}& A^2 \bigg[ \frac{3}{2}a_6 (S^2 - S_*^2)^2  + (3 a_6 S_*^2 + a_4) \notag \\
&\! \times (S^2 - S_*^2) + \frac{3}{2} a_6 S_*^4 + a_4 S_*^2 + \frac{a_2}{2} \bigg] \, .
\end{align}

For binary configurations with the BH spins aligned with the orbital angular momentum we may write the magnitude of the total spin as 
\begin{equation}
S_* = |\alpha_1S_1 + \alpha_2S_2|\,,
\end{equation} where $\alpha_{i}= \cos{\theta_{i*}} = \pm 1$ discriminates between parallel ($\alpha_i = +1$) and antiparallel ($\alpha_i = -1$) alignment of $\vec{S}_{i*}$ with $\vec{L}$.
For instance, up-down corresponds to $\alpha_1 = -\alpha_2 = 1$.
Because $J$ and $\xi$ are constant on $t_{\rm pre}$ one has 
\begin{subequations}
\begin{align}
J \simeq J_* &= \left|L + \alpha_1S_1 + \alpha_2S_2\right|
\, , \\
\xi \simeq \xi_* &= \frac{1}{M^2} \left[ (1+q) \alpha_1 S_1 + \left( 1+\frac{1}{q} \right) \alpha_2 {S}_2 \right]
\, ,
\end{align}
\end{subequations}
which implies that
\begin{align}
\frac{3}{2} a_6 S_*^4 + a_4 S_*^2 + \frac{a_2}{2}=0 \, .
\end{align}
Therefore, to leading order $\mathcal{O}(S^2-S_*^2)$ in the perturbation (i.e., assuming small misalignments between the BH spins and the orbital angular momentum), the total spin magnitude $S$ of binary BHs with nearly aligned spins satisfies
\begin{align}
\label{harmonic}
\frac{d^2}{dt^2}(S^2-S_*^2) + \omega^2 (S^2-S_*^2) \simeq 0 \, .
\end{align}
Equation~(\ref{harmonic}) has the form of a simple harmonic oscillator equation, where we identify the oscillation frequency 
\begin{align}
\omega = \sqrt{-A^2(3 a_6 S_*^2 + a_4)} \, .
\end{align}

The stability of the aligned spin configurations is determined by the sign of $\omega^2$:
\begin{itemize}
\item When $\omega^2 > 0$, Eq.~(\ref{harmonic}) describes simple harmonic oscillations in $S^2$ around $S_*^2$. The configuration is stable; small perturbations will cause precessional motion about the alignment.
\item When $\omega^2 = 0$, $S^2$ remains constant. This condition marks the onset of an instability.
\item When $\omega^2 < 0$, the oscillation frequency becomes complex, corresponding to an instability in the precessional motion leading to large misalignments of $\vec{S}_1$ and $\vec{S}_2$ with $\vec{L}$.
\end{itemize}
The points during the evolution of the binary BH at which the precession motion transitions from stable to unstable, or vice-versa, correspond to the solutions of $\omega^2 = 0$. Since $L$ (or equivalently $r$) is a monotonically decreasing function of time on the radiation-reaction timescale, such a point is a stable-to-unstable transition if $d\omega^2/dL > 0$ ($d\omega^2/dt < 0$) and an unstable-to-stable transition if $d\omega^2/dL < 0$ ($d\omega^2/dt > 0$).

The square of the oscillation frequency depends on $L$ according to
\begin{align}
\label{omega2}
\omega^2(L) ={}& \left[ L^2 - 2 \frac {q\alpha_1S_1 - \alpha_2S_2} {1-q} L + \left( \frac {q\alpha_1S_1 + \alpha_2S_2} {1-q} \right)^2 \right] \notag \\
&\! \times \left( L - \frac {q\alpha_1S_1 + \alpha_2S_2} {1+q} \right)^2 \left[\frac {3 M^{9} q^{5} (1-q)} {2 (1+q)^{11} L^7} \right]^2 \, .
\end{align}
It is clear from Eq.~(\ref{omega2}) that $\omega^2$ always has four roots, with two being the repeated root
\begin{align}
\label{L0repeated}
L_0 = \frac{q\alpha_1S_1 + \alpha_2S_2}{1+q} \, .
\end{align}
The corresponding value of the binary separation $r_0 = M^{-3}\eta^{-2}L_0^2$ always satisfies $r_0 \leq M$ and is thus unphysical. The other two roots are
\begin{align}
\label{Lpm}
L_\pm = \frac {q\alpha_1S_1 - \alpha_2S_2 \pm 2 \sqrt{-q\alpha_1\alpha_2S_1S_2}} {1-q} \, .
\end{align}
For $L_\pm$ to be real, we require that $\alpha_1\alpha_2 = -1$, leaving only the cases up-down and down-up.
If $\alpha_1 = - \alpha_2 = -1$ (down-up), then $L_\pm = - (\sqrt{qS_1} \pm \sqrt{S_2})^2 / (1-q)$ which is always nonpositive and can be discarded as unphysical. The only combination of $\alpha_1$ and $\alpha_2$ which makes $L_\pm$ both real and non-negative, thus indicating a physical precession instability, is $\alpha_1 = - \alpha_2 = 1$, which corresponds to the up-down configuration. Therefore, the up-up, down-down and down-up binary BH configurations are stable, whereas the up-down configuration can become unstable at separations where $\omega^2<0$. Any small misalignment of the BH spins with the orbital angular momentum leads to small oscillations of the spin vectors around the aligned configuration in the former three cases, but might cause large misalignments in the latter case.

In terms of only the parameters $M$, $q$, $\chi_1$ and $\chi_2$ of the BH binary, the expressions for the binary separations corresponding to the roots $L_\pm$ in the case of up-down spin alignment are
\begin{align} \label{rud}
r_{\mathrm{ud}\pm} = \frac{\left( \sqrt{\chi_1} \pm \sqrt{q\chi_2} \right)^4}{(1-q)^2} M
\, ,
\end{align}
which are precisely those derived in Ref.~\cite{2015PhRvL.115n1102G} by other means.
A third, alternative derivation is provided in Appendix~\ref{Near-alignment expansion}.

\begin{figure}[t]
\centering
\includegraphics[width=1.0\columnwidth]{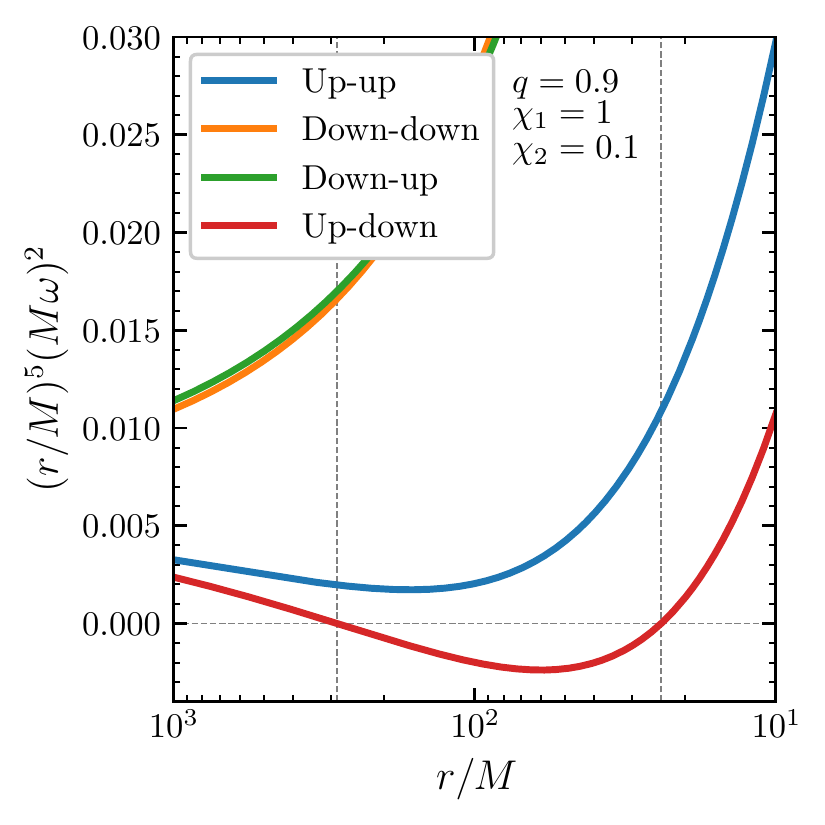}
\caption{Oscillation frequencies  for the four aligned configurations as a function of the binary separation $r$. The squared frequency $\omega^2$ is scaled by $r^5$ for clarity; see Eq.~(\ref{omegaud2}). The up-down configuration (red line) shows qualitatively different behavior to the other aligned configurations, with its oscillation frequency $\omega_\mathrm{ud}$ becoming complex (i.e. $\omega_\mathrm{ud}^2 < 0$) between $r_{\mathrm{ud}\pm}$ (dashed lines). In this example, the mass ratio is $q=0.9$, the dimensionless spins are $\chi_1=1.0$ and $\chi_2=0.1$, and the region of instability is given by $r_{\mathrm{ud}+} \simeq 285.6M$ and $r_{\mathrm{ud}-} \simeq 24.0M$.}
\label{fig omegasq}
\end{figure}

The oscillation frequency of the up-down configuration is given in terms of $r$ by
\begin{align}
\label{omegaud2}
M^2 \omega_\mathrm{ud}^2(r) ={}& \frac{9}{4} \left( \frac{1-q}{1+q} \right)^2 \left(\frac{M}{r}\right)^5 \left( 1 - \sqrt{r_{\mathrm{ud}0}/r} \right)^2 \notag \\
&\!\times \left( 1 - \sqrt{r_{\mathrm{ud}+}/r} \right) \left( 1 - \sqrt{r_{\mathrm{ud}-}/r} \right) \, ,
\end{align}
where
\begin{align}
r_{\mathrm{ud}0} =\left(\frac{\chi_1-q\chi_2}{1+q}\right)^2 M
\end{align} 
is the repeated root identified previously. One has
\begin{align}
\lim_{r/M\to \infty} M^2 \omega_{\mathrm{ud}}^2(r) = \frac{9}{4} \left( \frac{1-q}{1+q} \right)^2 \left(\frac{M}{r}\right)^5 > 0 \, ,
\end{align}
and hence the up-down configuration tends to stability at large orbital separations (past time infinity). Since $r_{\mathrm{ud}+} > r_{\mathrm{ud}-}$, the point $r = r_{\mathrm{ud}+}$ is a stable-to-unstable transition and $r = r_{\mathrm{ud}-}$ is an unstable-to-stable transition. In other words, $d\omega_\mathrm{ud}^2/dr|_{r_{\mathrm{ud}+}}>0$ and $d\omega_\mathrm{ud}^2/dr|_{r_{\mathrm{ud}-}}<0$. The up-down configuration is unstable for orbital separations $r_{\mathrm{ud}+} > r > r_{\mathrm{ud}-}$. An example of the behavior of $\omega^2$ is given in Fig. \ref{fig omegasq}.

In the equal-mass limit $q \to 1$, the precessional motion of up-down binaries tends to stability, since the time derivative of the total spin magnitude $S$ vanishes \cite{2017CQGra..34f4004G}. In the test-particle limit $q \to 0$, the behavior also tends to stability because $S\simeq S_1$ is constant.

\begin{figure}
\includegraphics[width=1.0\columnwidth]{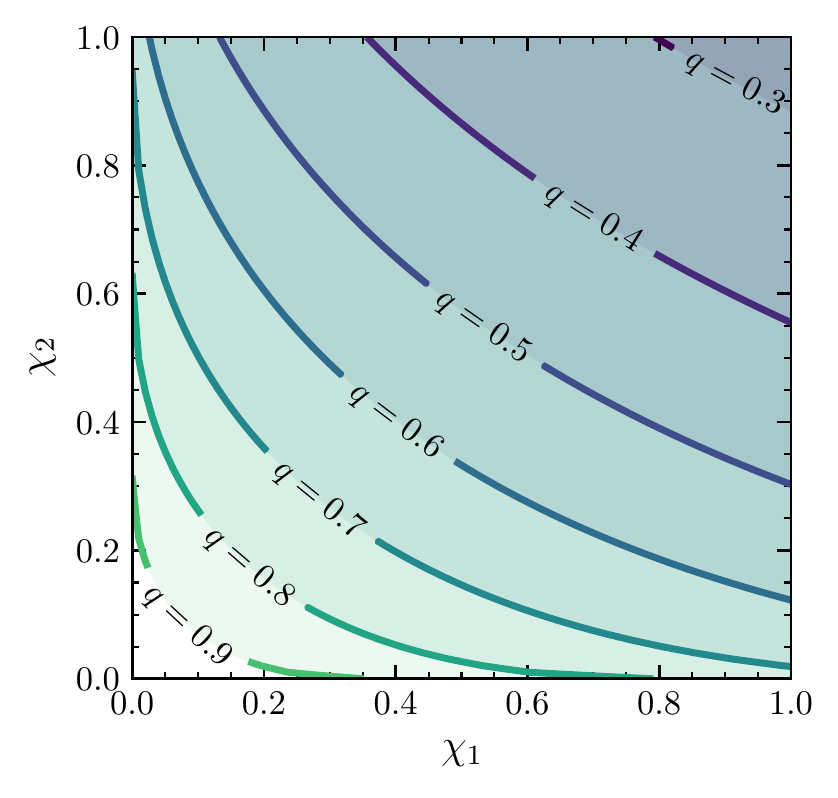}
\caption{Contours of constant mass ratio $q$ for values of the dimensionless spins $\chi_i$ that result in an instability threshold $r_{\mathrm{ud}+} = 10M$. Above each curve is the region of parameter space in which an up-down binary will experience the precessional instability at an orbital separation $r > 10M$. Below the curves, the instability takes place later in the inspiral where our PN approach is not valid.}
\label{fig q_contour}
\end{figure}

For an up-down binary to undergo the precessional instability, its parameters $q$, $\chi_1$, and $\chi_2$ must be such that the resulting instability onset satisfies $r_{\mathrm{ud}+} > 10M$, as this threshold represents the breakdown of the PN approximation \cite{2006PhRvD..74j4005B,2009PhRvD..79h4010C, 2009PhRvD..80h4043B}.
Figure~\ref{fig q_contour} shows contours in the  $\chi_1-\chi_2$ plane for various values of $q$ where $r_{\mathrm{ud}+} = 10M$.
For mass ratios close to unity, binaries with smaller dimensionless spins still result in a physical ($r_{\mathrm{ud}+} > 10M$) onset of instability. As the mass ratio becomes more extreme ($q \to 0$), only  binaries with $\chi_i \sim 1$ are affected by the instability, though much later in the inspiral.

\subsection{Numerical verification of the instability} \label{Numerical verification of the instability}

The analysis of Sec.~\ref{Binary BH spins as harmonic oscillators} is valid up to the onset of the precessional instability at the value of the binary separation $r = r_{\mathrm{ud}+}$, at which point spin precession invalidates the approximation of small misalignments between the BH spins and the orbital angular momentum. We therefore verify the existence of the instability with evolutions of binary BH spins performed via direct numerical integrations of the orbit-averaged spin precession equations.
The integrations are performed using the \textsc{python} module \textsc{precession}~\cite{2016PhRvD..93l4066G}.

\begin{figure*}[t]
\centering
\includegraphics[width=1.0\textwidth]{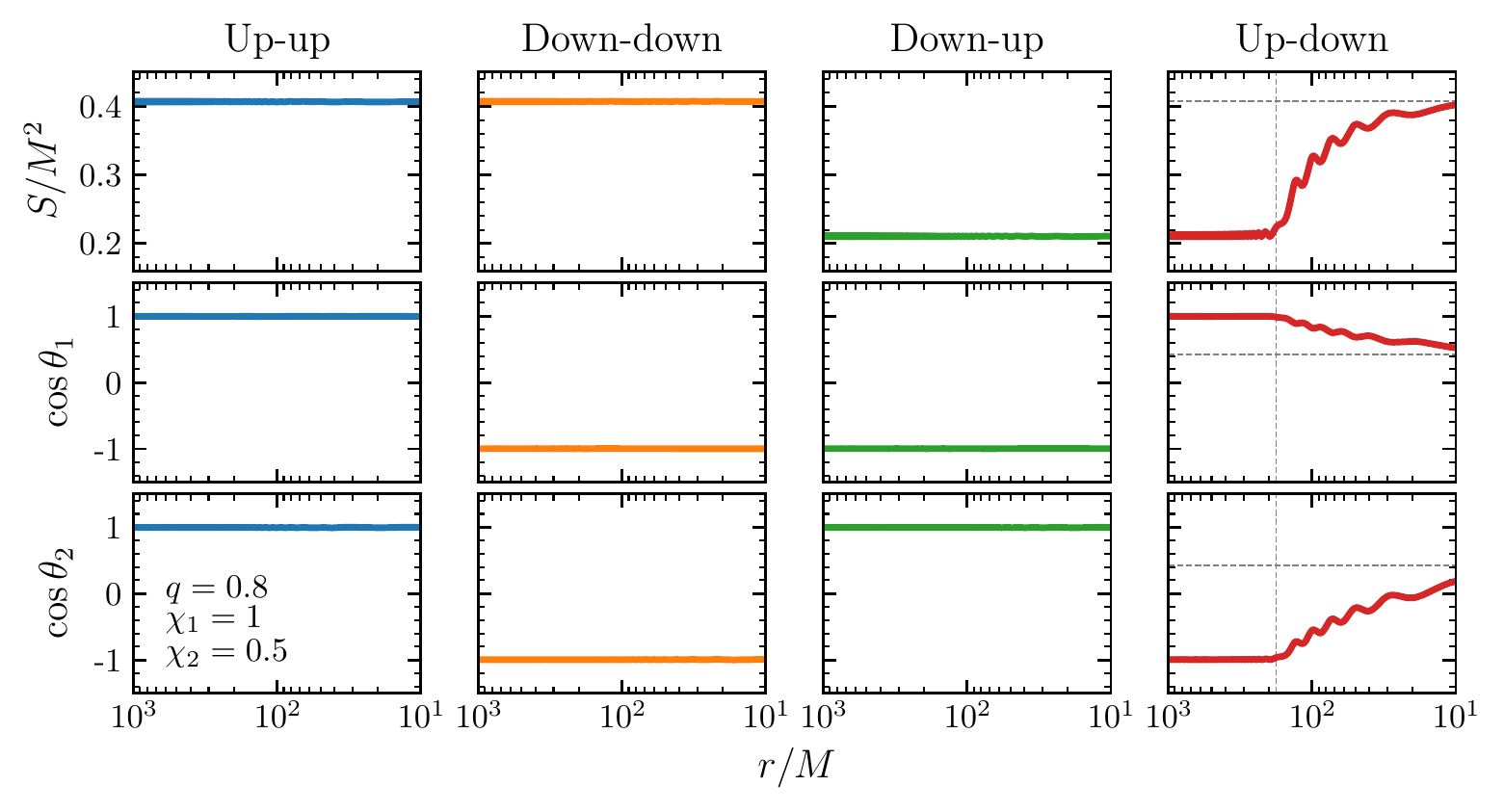}
\caption{Numerical evolutions of the total spin magnitude $S$ and misalignment angles $\cos\theta_i$ of four binary BHs with parameters $q=0.8$, $\chi_1=1.0$ and $\chi_2=0.5$, starting from an initial separation $r=1000M$ and ending at $r=10M$. Each panel shows a binary initially in a configuration with aligned spins up to a small perturbation of $5\degree$ in the angles $\theta_i$. The vertical dashed line in the right-most panel (up-down) shows the location of the instability onset $r_{\mathrm{ud}+} \simeq 177.5M$. The horizontal dashed lines mark the formal endpoint of the up-down instability obtained in the $r/M\to 0$ limit (Sec.~\ref{Instability limit}).}
\label{fig single}
\end{figure*}

The binaries are evolved from an initial separation $r=1000M$ down to a final separation $r=10M$.
The integrations are initialized by setting $\theta_1$, $\theta_2$, and $\Delta\Phi$ (or equivalently $\xi$, $J$ and $S$) at the initial separation.
The initial value of $\Delta\Phi$ is irrelevant (for these evolutions it was set to $\pi/2$).
We introduce an initial perturbation to each configuration by setting the initial values of $\theta_i$ to be $5\degree$ from the aligned configuration.
A number of binary BHs with varied mass ratios and dimensionless spins were evolved in this way to verify the existence of the instability.
As an example, the evolution of four binaries, one in each of the aligned spin configurations, with $q=0.8$, $\chi_1=1.0$ and $\chi_2=0.5$ is displayed in Fig.~\ref{fig single}.

In the exactly-aligned configurations each of $\theta_i$ and $S$ is constant, since such configurations are equilibrium solutions of Eqs.~(\ref{dS1/dt}-\ref{dL/dt}).
In the absence of the precessional instability, a small perturbation to $\theta_1$ and/or $\theta_2$ causes small amplitude oscillations around the equilibrium solutions.
For a perturbation in the angles as small as $5\degree$, a binary acts essentially as it would in the equilibrium configurations, as seen in the first three panels of Fig. \ref{fig single}: the angles $\theta_i$ remain approximately fixed at their initial values.
For the configurations in which the two BH spin vectors have the same alignment as each other with respect to $\vec{L}$ (up-up and down-down), the total spin magnitude remains at the initial value $S \simeq S_1 + S_2$.
In the down-up configuration, the total spin magnitude remains at the initial value $S \simeq |S_1 - S_2|$.
However, as is clear in the rightmost panel of Fig. \ref{fig single}, in the up-down configuration the values of $S$ and $\theta_i$ are not constant.
Though initially $S \simeq |S_1-S_2|$ and $\cos{\theta_1} = -\cos{\theta_2} \simeq 1$, after reaching the onset of the instability at $r = r_{\mathrm{ud}+} \simeq 177.5M$ the precessional motion moves the binary away from the initial up-down configuration.

\begin{figure*}[t] 
\centering
\includegraphics[width=1.0\textwidth]{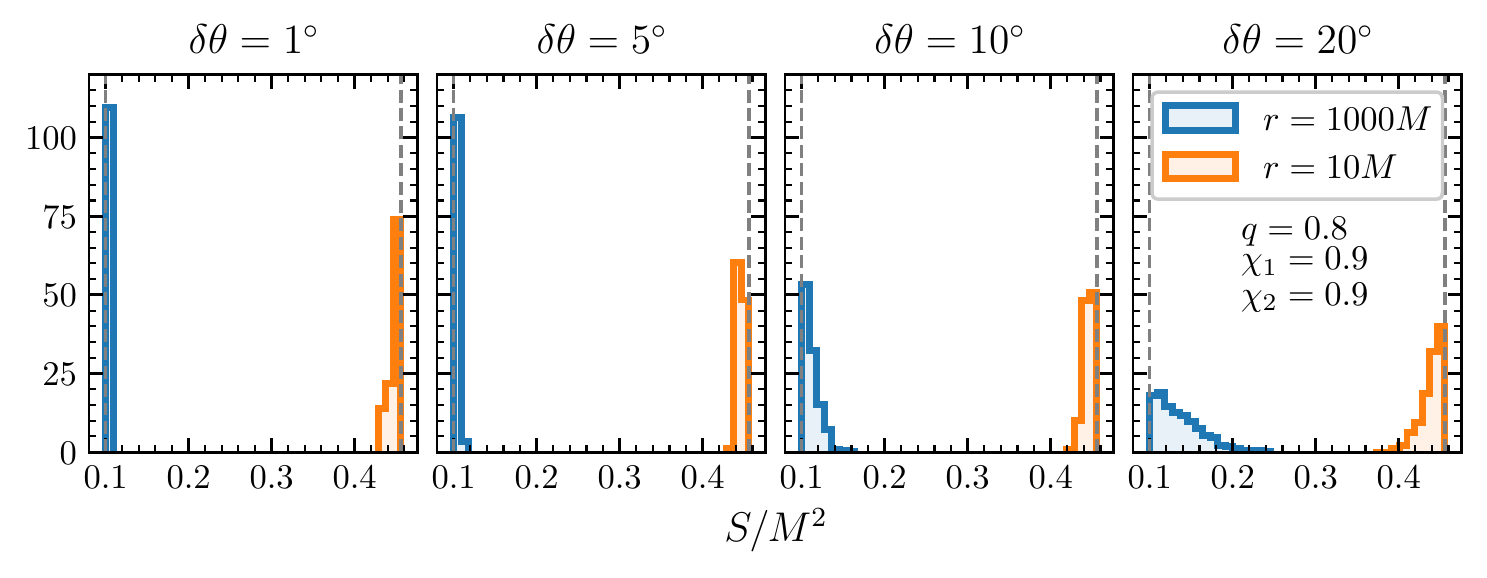}
\caption{The response of the up-down instability to different initial perturbations. Each panel shows a set of 1000 orbit-averaged evolutions. Binaries are initialized at $r=1000M$ with misalignments extracted from truncated Gaussians centered on the up-down configuration with widths $\delta\theta = 1\degree, 5\degree, 10\degree, 20\degree$, increasing progressively from the left to the right panel. Blue (orange) histograms show the corresponding values of the total spin $S$ at $r=1000M$ ($r=10M$). In this example we fix $q=0.8$ and $\chi_1=\chi_2=0.9$. Vertical dashed lines at $S=|S_1 \pm S_2|$ mark the asymptotic locations of up-down binaries before and after the instability. An animated version of this figure is available at \href{https://davidegerosa.com/spinprecession/}{www.davidegerosa.com/spinprecession}.}
\label{fig misalignments}
\end{figure*}

In Fig.~\ref{fig misalignments} we test the response of the up-down instability to the amplitude of the initial perturbation.
We evolve samples of binaries from $r=1000M$ to $r=10M$ and show their values of $S$ at both the initial and the final separations. Binaries are initialized by extracting the misalignments from half-Gaussian distributions in $\cos\theta_i$ ($i=1,2$) with widths $1 - \cos\delta\theta$ centred on the exact up-down configuration, where $\delta\theta = 1\degree, 5\degree, 10\degree, 20\degree$. The initial value of $\Delta\Phi$ is irrelevant and is here extracted uniformly in $[-\pi,\pi]$. In this example we fix $q=0.8$, $\chi_1=\chi_2=0.9$. 

Our numerical evolutions show a somewhat surprising result: \emph{binaries do not tend to disperse in parameter space as one would expect from an instability, but present a well-defined endpoint.} This effect is sharper for binaries very close to up-down. Increasing the initial misalignment $\delta\theta$ dilutes both the initial and the final spin distributions, although the same trend remains present up to $\delta\theta \lesssim 20\degree$. Binaries that undergo the up-down instability at some large separation are likely to be found in a different, but very specific region of the parameter space at the end of the inspiral. We now aim to find this location analytically.

\section{Resonant configurations} \label{Resonant configurations}

Spin-orbit resonances \cite{2004PhRvD..70l4020S} are special configurations where the three vectors $\vec{L}$, $\vec{S}_1$, and $\vec{S}_2$ are coplanar and jointly precess about $\vec{J}$. There are two families of resonant solutions, defined by $\Delta\Phi=0$ and $\Delta\Phi=\pi$. 
The previous analysis of Ref.~\cite{2015PhRvL.115n1102G} indicated that the up-down configuration at separations $r > r_{\mathrm{ud}+}$ ($r < r_{\mathrm{ud}-}$) is  $\Delta\Phi = 0$ ($\Delta\Phi = \pi$) resonance.
The end-point of the up-down instability is thus deeply connected to the evolution of these special solutions.
As a building block to analyze the up-down configuration, in this section we present new advances toward understanding  spin-orbit resonances in a semianalytic fashion.

\subsection{Locating the resonances} \label{Locating resonances}

For fixed values of $q$, $\chi_1$, $\chi_2$, $J$, and $L$,  geometrical constraints restrict the allowed values of $S$ and $\xi$ to \cite{2015PhRvD..92f4016G}
\begin{subequations}
\begin{align} \label{S range}
S_\mathrm{min} \leq \ &S \leq S_\mathrm{max}
\, , \\
\label{effpotlimit}
\xi_-(S) \leq \ &\xi \leq \xi_+(S)
\, ,
\end{align}
\end{subequations}
where 
\begin{subequations} \label{Slim}
\begin{align}
\label{Smin}
S_\mathrm{min} ={}& \max \{|J-L| , |S_1-S_2|\} \, , \\
\label{Smax}
S_\mathrm{max} ={}& \min \{J+L , S_1+S_2\} \, ,\\
\label{xipm}
\xi_\pm ={}& \bigg\{ \frac{J^2-L^2-S^2}{4qM^2S^2L} \left[ (1+q)^2S^2 - (1-q^2)(S_1^2-S_2^2) \right] \notag\\
&\! \pm (1-q^2) {\left[ J^2-(L-S)^2 \right]^{1/2} \left[ (L+S)^2-J^2 \right]^{1/2}} \notag \\
&\! \times {\left[ S^2-(S_1-S_2)^2 \right]^{1/2} \left[ (S_1+S_2)^2-S^2 \right]^{1/2}} \bigg\} \, .
\end{align}
\end{subequations}
Together, the functions $\xi_\pm(S)$ form a closed convex loop in the $S-\xi$ plane, which implies that the inequalities (\ref{S range}-\ref{effpotlimit}) can be rewritten as 
\begin{align}
S_-\leq S \leq S_+ \, ,
\end{align}
where $S_\pm$ are the solutions of $\xi=\xi_\pm(S)$. One can trivially prove that the condition $\xi=\xi_\pm(S)$ is equivalent to either alignment ($\sin\theta_i=0$) or coplanarity ($\sin\Delta\Phi=0$). Generic spin precession can be described as a quasiperiodic motion of $S$ between the two solutions $S_\pm$.  Spin-orbit resonances correspond to the specific case where $S_-=S_+$, i.e. $d\xi_{\pm}/dS=0$. In this case, $S$ is constant: the three momenta are not just coplanar, but stay coplanar on the precession timescale $t_{\rm pre}$. As we will see later in Sec.~\ref{Evolution of resonances}, coplanarity is also preserved on the longer radiation-reaction timescale $t_{\rm rad}$.
 
The conditions $\xi=\xi_\pm(S)$ can be squared and cast into the convenient form
\begin{align}
\label{S2 cubic}
\Sigma(S^2) &\equiv \sigma_6 S^6 + \sigma_4 S^4 + \sigma_2 S^2 + \sigma_0 = 0 \, ,
\end{align}
where the coefficients $\sigma_i$ are real multiples of the $a_i$ in Eq.~(\ref{coeffa6}-\ref{coeffa0}) and are given explicitly in Appendix~\ref{longappendix}.

The existence of physical solutions can be characterized using the discriminant
\begin{align}
\label{cubic discriminant}
\Delta \equiv \sigma_4^2 \sigma_2^2 - 4 \sigma_6 \sigma_2^3 - 4 \sigma_4^3 \sigma_0 - 27 \sigma_6^2 \sigma_0^2 + 18 \sigma_6 \sigma_4 \sigma_2 \sigma_0
\, .
\end{align}
In particular:
\begin{itemize}
\item If $\Delta > 0$, then $\Sigma(S^2)$ has three distinct real roots. These are the physical solutions $S_-$ and $S_+$ identified in Ref.~\cite{2015PhRvD..92f4016G}, plus a spurious root that does not satisfy Eqs.~(\ref{S range}-\ref{effpotlimit}).
\item If $\Delta = 0$, the two solutions $S_-$ and $S_+$ coincide and correspond to a spin-orbit resonance.
\item If $\Delta < 0$, the polynomial $\Sigma(S^2)$ only admits one spurious real root, thus implying that the geometrical constraints in Eqs.~(\ref{S range}-\ref{effpotlimit}) cannot be satisfied for the assumed set of parameters $(q,\chi_1,\chi_2,J,\xi,L)$.
\end{itemize}
Therefore, physical spin precession takes place whenever $\Delta \geq 0$. The limiting case of the spin-orbit resonances can be located by solving $\Delta = 0$.

The discriminant reported in in Eq.~(\ref{cubic discriminant}) may be recast as a fifth-degree polynomial in $J^2$,
\begin{align}
\label{cubic discriminant J2}
\Delta (J^2) = \delta_{10} J^{10} + \delta_8 J^8 + \delta_6 J^6 +\delta_4 J^4 + \delta_2 J^2 + \delta_0 \, ,
\end{align}
where the coefficients $\delta_i$ are lengthy (but real and algebraic) expressions containing $q$, $S_1$, $S_2$, $\xi$, and $L$; see Appendix~\ref{longappendix}. In particular,
\begin{align}
\label{delta10}
\delta_{10} = - 4q^3(1-q)^2(1+q)^8L^2 \leq 0 \, .
\end{align}

\subsection{Number of resonances} \label{numberres}

Any fifth-degree polynomial has at most two bound intervals and one unbound interval in which it is positive. The two bounds intervals are the only possible locations in which spin precession can occur. We now prove that only one of these can be physical. 

To this end, it is useful to look at the asymptotic limit  $r\to \infty$. While $J$ diverges in this limit, one has \cite{2015PhRvD..92f4016G}
\begin{align}
\label{kappa}
\kappa_\infty \equiv \lim_{r/M\to \infty} {\vec S}\cdot \hat{\vec L}= \lim_{r/M\to \infty}  \frac{J^2-L^2}{2L} =\mathrm{constant} \, .
\end{align}
The constraints $|\cos\theta_1|\leq 1$ and $|\cos\theta_2|\leq1$ can be translated into
\begin{align}
\label{kappamin}
\kappa_{\infty} &\geq \max \bigg\{ \frac{M^2\xi - (q^{-1}-q)S_1}{1+q^{-1}},
\frac{M^2\xi - (q^{-1}-q)S_2}{1+q} \bigg\} \, , \\
\label{kappamax}
\kappa_{\infty} &\leq \min \bigg\{ \frac{M^2\xi + (q^{-1}-q)S_1}{1+q^{-1}},
\frac{M^2\xi + (q^{-1}-q)S_2}{1+q} \bigg\} \, .
\end{align}
Therefore,  the support of $(J^2-L^2)/2L$ (hence $J$) is a single bounded interval at large separations: only one range of $J$ is allowed and is it bounded by two resonances. Proving by contradiction, let us now assume that the support of $J$ does \emph{not} remain a single interval. A bifurcation would be present at some finite separation where the number of valid ranges goes from one to two. At this bifurcation point, two different values of $dJ/dr$ must coexist for the same values of $q$, $\chi_1$, $\chi_2$, $\xi$, $J$. This is only possible if the two configurations have different values of $S$. However, at the bifurcation point one necessarily has $\Sigma(S^2)=0$ and thus only one value of $S$ is allowed.

Our proof is  consistent with the extensive numerical exploration presented in  Refs.~\cite{2015PhRvD..92f4016G,2015PhRvL.114h1103K}: there are always two spin-orbit resonances for any values of $q$, $\chi_1$, $\chi_2$, $\xi$, and $r$. The two resonances are characterized by $\Delta\Phi = 0$ and $\Delta\Phi = \pi$. In particular, the $\Delta\Phi=0$ ($\Delta\Phi=\pi$) resonance corresponds to the maximum (minimum) value of $J$, i.e.,
\begin{align}
\label{Jextremes}
J^{(\Delta\Phi=\pi)}\leq J \leq J^{(\Delta\Phi=0)} \, . 
\end{align}

\begin{figure}[t] 
\centering
\includegraphics[width=1.0\columnwidth]{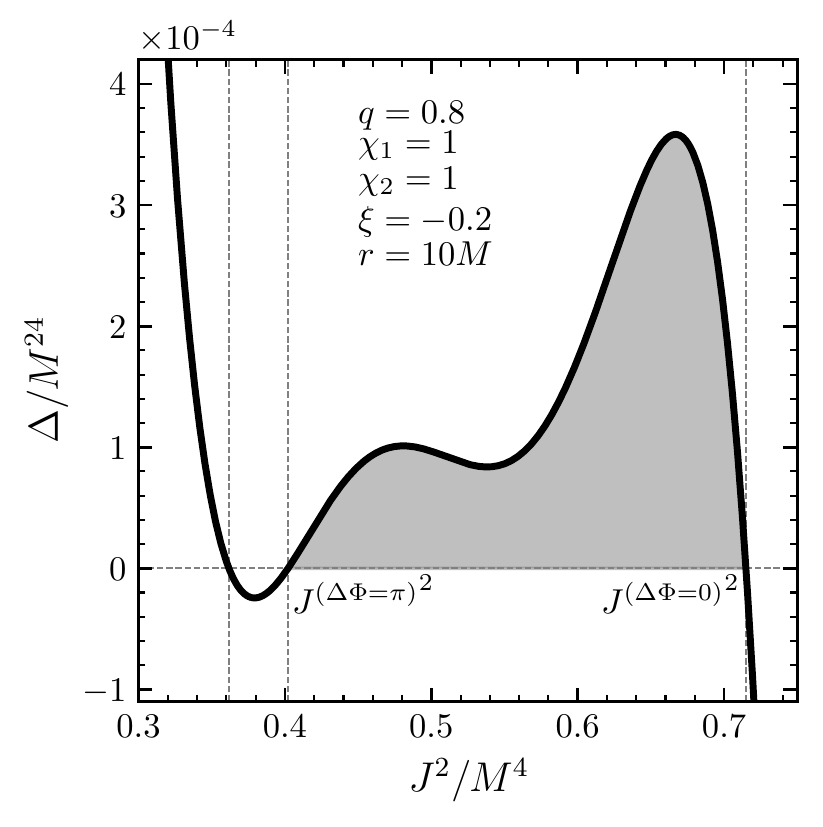}
\caption{The discriminant $\Delta$ of the third-degree polynomial $\Sigma$ as a function of $J^2$ for a binary BH at a separation $r=10M$ with mass ratio $q=0.8$, dimensionless spins $\chi_1=\chi_2=1$ and effective spin $\xi=-0.2$. The discriminant has three real roots (vertical dashed lines). The shaded area between the two roots in which $\Delta \geq 0$ corresponds to the region in which physical spin precession takes place. This region is bounded on the right (left) by the $\Delta\Phi = 0$ ($\Delta\Phi = \pi$) resonance.}
\label{fig discriminant}
\end{figure}

An example is shown in Fig.~\ref{fig discriminant}. The  region of $J^2$ where physical spin precession takes place is characterized by $\Delta(J^2)>0$. The spin-orbit resonances correspond to two of the roots of $\Delta(J^2)=0$.

\subsection{Evolution of resonances} \label{Evolution of resonances}

Next, we prove that a binary in a resonant configuration remains  resonant under radiation reaction. 

Let us label two binaries $A$ and $B$. The binaries share the same values of the radiation-reaction constants of motions $q, \chi_1,\chi_2$, and $\xi$. Suppose binary $A$ is a $\Delta\Phi=0$ resonance at separation $r$ and binary $B$ is  a $\Delta\Phi=0$ resonance at $r+\delta r$.  Again by contradiction, let us now assume that $A$ and $B$ do not coincide. From Eq.~(\ref{Jextremes}) one has $J_A(r)>J_B(r)$ and  $J_A(r+\delta r)<J_B(r+\delta r)$. 
At some location $r<\tilde r< r+\delta r$ one must have $J_A(\tilde r) = J_B(\tilde r)$, but $d J_{A}/dr|_{\tilde r} \neq d J_{B}/dr|_{\tilde r}$. In other terms, the inspiral trajectory of the two binaries must cross in the $J\!-\!r$ plane. This is possible only if the two binaries have different values of $S$ at $\tilde r$, i.e. $S_A(\tilde r)\neq S_B(\tilde r)$. Taking the limit  $\delta r/M\to 0$, the location of the crossing point  can be made arbitrarily close to the initial separation $r$. At this location, $J_A=J_B$ identifies a resonance, where only one value of $S$ is allowed. It follows that the two binaries $A$ and $B$ must coincide. An analogous proof can be carried out for $\Delta\Phi=\pi$.

\subsection{Resonance asymptotes} \label{Resonance asymptotes}

Further progress can be made by studying the dynamics of resonant configurations at infinitesimal separations $r/M \to 0$ (or equivalently $L/M^2\to 0$). Although unphysical, this limit provides the asymptotic conditions of our PN evolutions.

Let us denote the effective spin of the up-up and up-down configuration with respectively
\begin{subequations}
\begin{align}
\label{xi_uu}
\xi_\mathrm{uu} = \frac{1}{M^2} \left[ \left(1+q\right)S_1 + \left(1 + \frac{1}{q}\right)S_2 \right] \, , \\
\label{xi_ud}
\xi_\mathrm{ud} = \frac{1}{M^2} \left[ \left(1+q\right)S_1 - \left(1 + \frac{1}{q}\right)S_2 \right] \, .
\end{align}
\end{subequations}
As $r/M \to 0$, one has that $J \to S$ and $\Delta$ is increasingly dominated by the term with the least power of $L$. In particular, one gets
\begin{align}
\lim_{L/M^2 \to 0} \frac{\Delta}{\delta_{10}} = \prod_{j=1}^{5} (S^2 - \lambda_j) \, .
\end{align}
The roots $\lambda_i$ of this expression are given by
\begin{subequations}
\begin{align}
\lambda_1 &= \lambda_2 = \frac{(1-q)(qS_1^2-S_2^2)}{q} \, , \\
\lambda_3 &= \frac{(1-q)(qS_1^2-S_2^2)}{q} + \frac{M^4q\xi^2}{(1+q)^2} \, , \\
\lambda_4 &= (S_1-S_2)^2 \, , \\
\lambda_5 &= (S_1+S_2)^2 \, .
\end{align}
\end{subequations}
The constraint $|\xi| \leq \xi_{\rm uu}$ implies the following series of inequalities:
\begin{align}
\lambda_1 = \lambda_2 \leq \min\{\lambda_3,\lambda_4\} \leq \max\{\lambda_3,\lambda_4\} \leq \lambda_5 \, ,
\end{align}
with
\begin{align}
\max\{\lambda_3,\lambda_4\} =
\begin{cases}
\lambda_4 \quad \mathrm{if} \quad |\xi| \leq |\xi_{\rm ud}|
\, , \\
\lambda_3 \quad \mathrm{if} \quad |\xi| > |\xi_{\rm ud}|
\, .
\end{cases}
\end{align}
Since $\Delta \leq 0$ as $J \to +\infty$ [cf. Eq.~(\ref{delta10})], the two bounded intervals of $J^2$ in which $\Delta \geq 0$ are $[\lambda_1, \min\{\lambda_3,\lambda_4\}]$ and $[\max\{\lambda_3,\lambda_4\}, \lambda_5]$. 
Furthermore, in this limit Eq.~(\ref{S range}) reduces to 
\begin{equation}
\lambda_4 \leq S^2 \leq \lambda_5\,,
\end{equation}
which implies that the single physical interval in which spin precession takes places is given by
\begin{align}
S^2 \in [\max\{\lambda_3,\lambda_4\}, \lambda_5]\,.
\end{align}
The boundaries $S = \max\{\sqrt{\lambda_3},\sqrt{\lambda_4}\}$ and $S = \sqrt{\lambda_5}$ of this region identify the asymptotic locations of the $\Delta\Phi=\pi$ and $\Delta\Phi=0$ resonances, respectively. Thus, the value $S^{(\Delta\Phi = 0)}$ of S in the $\Delta\Phi = 0$ spin-orbit resonance asymptotes to
\begin{align} \label{asymptote 0}
\lim_{r/M \to 0} S^{(\Delta\Phi = 0)} = S_1 + S_2
\, ,
\end{align}
and the value $S^{(\Delta\Phi = \pi)}$ of S in the $\Delta\Phi = \pi$ resonance asymptotes to
\begin{align} \label{asymptote pi}
\lim_{r/M \to 0} S^{(\Delta\Phi = \pi)} =
\begin{cases}
|S_1 - S_2| \quad\mathrm{if} \quad |\xi| \leq |\xi_{\rm ud}|\, , \\
\\
\sqrt{\displaystyle\frac{(1-q)(qS_1^2-S_2^2)}{q} + \frac{M^4q\xi^2}{(1+q)^2}}
\\
\qquad  \mathrm{if} \quad |\xi| > |\xi_{\rm ud}|
\, .
\end{cases}
\end{align}

The corresponding values of the misalignment angles $\theta_i$ are found by imposing the coplanarity condition $\sin(\Delta\Phi)=0$ that characterizes the resonances. This yields
\begin{equation}
\sin^2\theta_1 \sin^2\theta_2 = \left(\frac{S^2-S_1^2-S_2}{2 S_1 S_2} - \cos\theta_1 \cos\theta_2\right)^2 \, ,
\end{equation}
which can be solved together with Eq.~(\ref{xi}) to find $\cos\theta_1$ and $\cos\theta_2$. For the $\Delta\Phi=0$ resonance one gets
\begin{align}
\label{thetasphi0lim}
\lim_{r/M \to 0} \cos\theta_1^{(\Delta\Phi=0)} = 
\lim_{r/M \to 0} \cos\theta_2^{(\Delta\Phi=0)} = \frac{\xi}{\xi_{\rm uu}}
\, .
\end{align}
In words, the two spins tend to be equally misaligned with $\vec L$ but coaligned with each other. Hints of this trend had been reported in Refs.~\cite{2004PhRvD..70l4020S,2010PhRvD..81h4054K,2013PhRvD..87j4028G}. For $\Delta\Phi=\pi$, the angles asymptote to 
\begin{subequations}
\begin{align}
\label{ct1pi}
\lim_{r/M \to 0} \cos\theta_1^{(\Delta\Phi = \pi)} &=
\begin{cases}
\displaystyle\frac {\xi} {\xi_{\rm ud}}
 \quad\mathrm{if} \quad |\xi| \leq |\xi_{\rm ud}|\, , \\
 \\
\displaystyle \frac{ \xi^2 + \xi_{\rm uu} \xi_{\rm ud} }{2(1+q) S_1 \xi}M^2
 \\
 \qquad\mathrm{if} \quad |\xi| > |\xi_{\rm ud}|\,,
\end{cases}
\\
\lim_{r/M \to 0} \cos\theta_2^{(\Delta\Phi = \pi)} &=
\begin{cases}
\displaystyle -\frac{\xi}{\xi_{\rm ud}}
 \quad\mathrm{if} \quad |\xi| \leq |\xi_{\rm ud}|\, , \\
 \\
\displaystyle \frac{q( \xi^2  -\xi_{\rm uu} \xi_{\rm ud})}{2(1+q) S_2 \xi}M^2
 \\
 \qquad\mathrm{if} \quad |\xi| > |\xi_{\rm ud}|\,.
\end{cases}
\label{ct2pi}
\end{align}
\end{subequations}

\begin{figure}[t] 
\centering
\includegraphics[width=1.0\columnwidth]{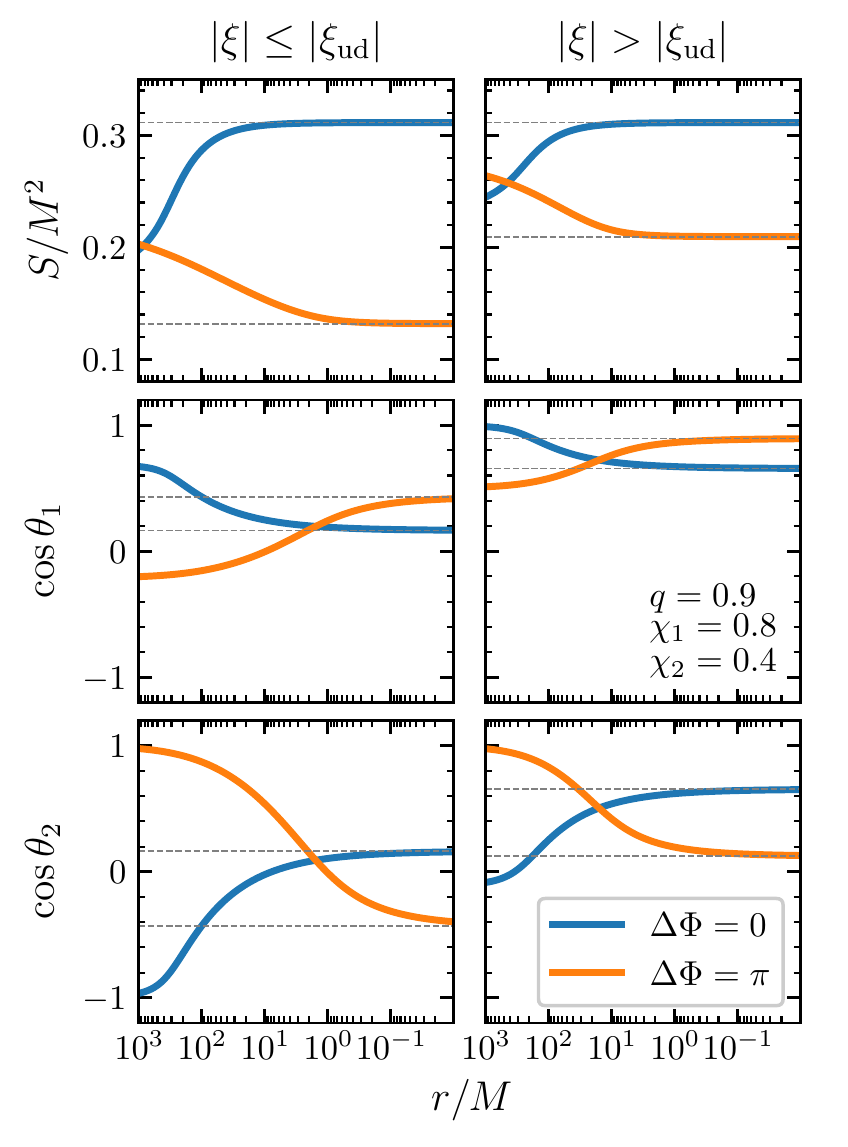}
\caption{Evolutions of resonant configurations to small separations. Top, middle, and bottom panels show $S/M^2$, $\cos\theta_1$, and $\cos\theta_2$, respectively. Blue (orange) curves correspond to resonances with $\Delta=0$ ($\Delta=\pi$), while binaries in the left (right) panel satisfy $|\xi|\leq |\xi_{\rm ud}|$ ($|\xi|> |\xi_{\rm ud}|$). We fix $q=0.9$, $\chi_1=0.8$, $\chi_2=0.4$, $\xi=0.1$ (left), $0.4$  (right).  For this set of parameters one has $\xi_{\rm ud}\simeq 0.23$ and $\xi_{\rm uu}\simeq 0.61$. Dashed gray lines indicate the $r/M\to 0$ limits predicted in Sec.~\ref{Resonance asymptotes}. The region $r\lesssim 10M$ should be considered unphysical but is included to test our analytical calculations.}
\label{fig res_evolve}
\end{figure}

Figure~\ref{fig res_evolve} shows the evolution of four resonant configurations for $\Delta\Phi=0,\pi$ and the two cases $|\xi| \leq |\xi_{\rm ud}|$ and  $|\xi| > |\xi_{\rm ud}|$. At each separation we locate the roots of $\Sigma(S^2)=0$ numerically using the algorithm implemented in the \textsc{precession} code~\cite{2016PhRvD..93l4066G}. Because resonant binaries remain resonant during the inspiral (Sec.~\ref{Evolution of resonances}), those curves also correspond to individual evolutions. As $r/M \to 0$,  binaries asymptote to the limits predicted above.

\section{Up-down endpoint} \label{Up-down endpoint}

\subsection{Instability limit} \label{Instability limit}

The analysis of Sec.~\ref{Resonant configurations}  allows us to find the asymptotic endpoint of the up-down configuration.
As first shown in Ref.~\cite{2015PhRvL.115n1102G}, the up-down configuration is a $\Delta\Phi=0$ resonance  for $r > r_{\mathrm{ud}+}$. This can be immediately seen using the expressions in Sec.~\ref{numberres}. As $r/M\ \to \infty$, the up-down configuration corresponds to $\kappa_\infty = S_1-S_2$ which maximizes the allowed range of $\kappa_\infty$ given in Eqs.~(\ref{kappamin}-\ref{kappamax}), and hence that of $J$. The largest value of $J$ for a given $\xi$ corresponds to the $\Delta\Phi=0$ resonance [cf. Eq.~(\ref{Jextremes})].

A binary which is arbitrarily close to up-down before the instability onset, therefore, will be arbitrarily close to a $\Delta\Phi=0$ spin-orbit resonance. As shown in Sec.~\ref{Evolution of resonances}, resonant binaries remain resonant during the entire inspiral. The formal $r/M\to 0 $ limit of the up-down instability is that of a $\Delta\Phi=0$ resonance with the correct value of the effective spin. This can be obtained directly from Eqs.~(\ref{asymptote 0}) and Eq.~(\ref{thetasphi0lim}) by setting $\xi=\xi_{\rm ud}$. 

The key result of this paper is that the endpoint of the up-down instability consists of a binary configuration with
\begin{align} \label{endpointanalytic}
\cos\theta_1= \cos\theta_2 = \frac{\chi_1-q\chi_2}{\chi_1+q\chi_2}\quad\; {\rm and} \quad\; \Delta\Phi=0
\, ,
\end{align}
which is equivalent to Eq.~(\ref{endpointintro}).
Up-down binaries start their inspiral with $S=|S_1-S_2|$ and asymptote to $S= S_1+S_2$ as given by Eq.~(\ref{asymptote 0}), thus spanning the entire range of available values of $S$, cf. Eq.~(\ref{S range}). 

An example is reported in Fig.~\ref{fig single}. Despite being obtained for $r/M\to 0$, the spin configuration in   Eq.~(\ref{endpointanalytic}) well describes the inspiral endpoint. Similarly, Fig.~\ref{fig misalignments} shows that binaries initially close to the up-down configuration all evolve to this precise location in parameter space.

\begin{figure}[t] 
\centering
\includegraphics[width=0.85\columnwidth]{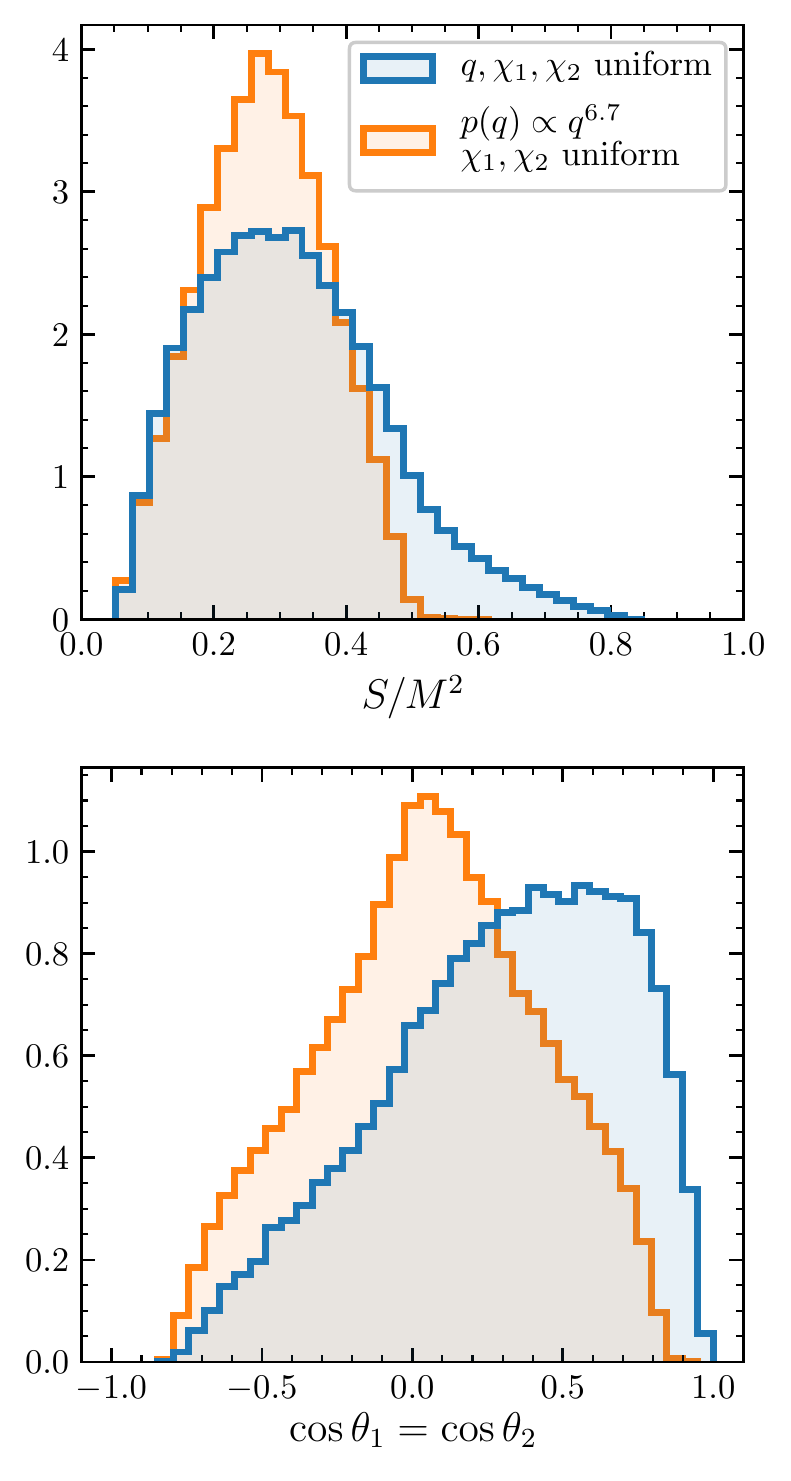}
\caption{Analytic up-down endpoint distribution in the $r/M\to 0$ limit. Top and bottom panels show the endpoint distributions of $S$ and $\cos\theta_1=\cos\theta_2$. Blue histograms are obtained distributing $q$ uniformly; orange histograms  assume $p(q)\propto q^{6.7}$ as observed by LIGO/Virgo \cite{2019ApJ...882L..24A}. In both cases, we distribute $\chi_1$ and $\chi_2$ uniformly  and assume $q,\chi_1,\chi_2\in  [0.1,1]$.}
\label{fig analytic_endpoint}
\end{figure}

Figure~\ref{fig analytic_endpoint} illustrates the formal $r/M\to 0$ distribution for two simple BH populations. In particular, we distribute mass ratios $q$ either uniformly or according to the astrophysical population inferred from the first GW events, $p(q)\propto q^{6.7}$ (cf. Model B in Ref.~\cite{2019ApJ...882L..24A}; see also Ref.~\cite{2020ApJ...891L..27F}). In both cases, we take $q\in[0.1,1]$ and assume spin magnitudes $\chi_i$ are distributed uniformly in $[0.1,1]$. The LIGO/Virgo-motivated population strongly favors equal mass events. For $q \simeq 1$ the instability endpoint is given by $S/M^2 \simeq (\chi_1+\chi_2)/4$ and $\cos\theta_i \simeq (\chi_1-\chi_2)/(\chi_1+\chi_2)$, which implies that the corresponding distributions are peaked at $S/M^2\simeq (0.1+1)/4=0.275$ and $\cos\theta_i\simeq 0$. If $q$ differs from unity, the endpoint values of both $S$ and $\cos\theta_i$ are, on average, larger. For the case where mass ratios are drawn uniformly, unequal-mass binaries populate the region of Fig.~\ref{fig analytic_endpoint} with $S/M^2\gtrsim 0.5$ and $\cos\theta_i\gtrsim 0.7$.

As a mathematical curiosity, we note that if one places a binary in the up-down configuration at $r<r_{\rm ud-}$, this must necessarily be a $\Delta\Phi=\pi$ resonance (cf. Ref.~\cite{2015PhRvL.115n1102G}). Indeed, for $\xi=\xi_{\rm ud}$ Eqs.~(\ref{ct1pi}-\ref{ct2pi}) return  $\cos\theta_1^{(\Delta\Phi = \pi)} = - \cos\theta_2^{(\Delta\Phi = \pi)} =1$.  We stress that this case is not physically relevant. Before reaching $r_{\rm ud-}$, binaries have already reached $r_{\rm ud+}$ and thus left the up-down configuration. Unless $q$ is very close to unity and $\chi_2$ is very close to zero, the separations $r_{\rm ud-}$ is typically smaller than $10 M$ (or even $1M$): it is hard, if not impossible, to conceive plausible astrophysical mechanisms that can place binaries in the up-down configuration so close to merger.

\subsection{Stability-to-instability transition} \label{Instability transition}

\begin{figure}[t] 
\centering
\includegraphics[width=1.0\columnwidth]{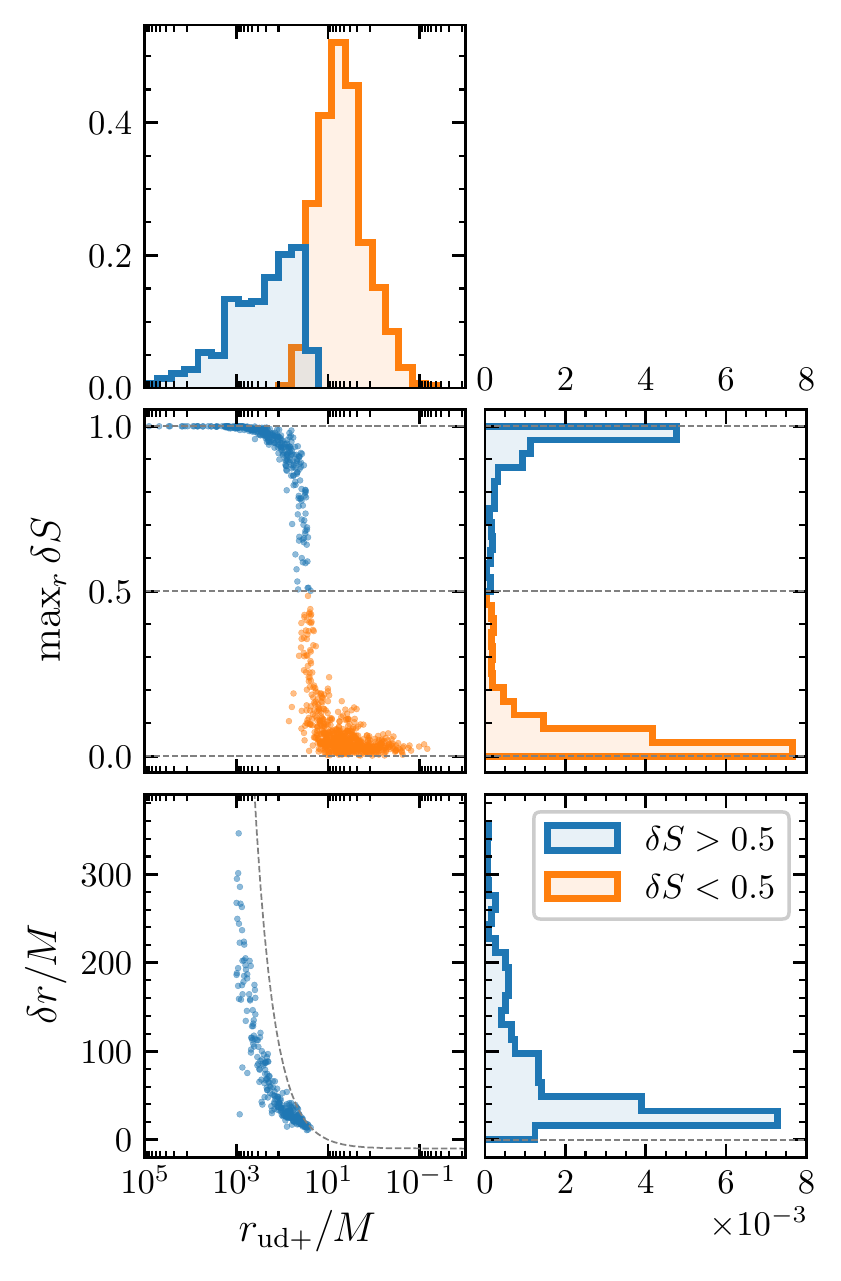}
\caption{Distribution of $\max_r \delta S$ and $\delta r$ as function of the instability onset $r_{\rm ud+}$ for a statistical sample of 1000 up-down binaries. We distribute $q$, $\chi_1$, and $\chi_2$ uniformly in $[0.1,1]$, evolve from $r_{i}=1000M$ to $r_{\rm f}=10M$, and initialize the spin misalignments from Gaussian distributions with widths $\delta\theta = 10\degree$ centered on the up-down configuration. The blue (orange) subpopulation indicates sources that do (not) reach $\delta S = 0.5$ by the end of the inspiral. By definition, $\delta r$ can only be computed for the subpopulation with $\max_r \delta S > 0.5$, with a minimum value of $\delta r \leq r_{\mathrm{ud}+} - r_\mathrm{f}$ (dashed line).}
\label{fig deltaS}
\end{figure}

During the inspiral, unstable up-down binaries evolve from $S=|S_1-S_2|$ to  $S= S_1+S_2$. The transition between the two values can only start after binaries enters the instability regime ($r<r_{\rm ud +}$) and is halted by the merger (or, to be more conservative, by the PN breakdown). To quantify the transition properties, it is useful to define the parameter
\begin{align}
\delta S \equiv \frac {S - |S_1-S_2|} {(S_1+S_2) - |S_1-S_2|} \, ,
\end{align}
such that $\delta S=0$ corresponds to stability and $\delta S=1$ corresponds to the formal $r/M\to 0$ endpoint.

Figure~\ref{fig deltaS} shows the distribution of $\delta S$ and $r_{\rm ud+}$ resulting from numerical integrations of up-down binaries. We distribute $q$, $\chi_1$, and $\chi_2$ uniformly in $[0.1,1]$ and evolve from $r_{i}=1000M$ to $r_{\rm f}=10M$. Binaries with $r_{\rm ud+}<r_{\rm i}$ are initialized as up-down and might become unstable during the integration. Binaries with $r_{\rm ud+}>r_{\rm i}$, on the other hand, are already unstable at the start of our integrations. We therefore initialized them as $\Delta\Phi=0$ resonances at $r=r_{\rm i}$. In both cases, we introduce a misalignment perturbation $\delta\theta = 10\degree$ following the same procedure of Sec~\ref{Numerical verification of the instability}.

We consider the largest value of $\delta S$ reached between $r_{\rm i}$ and $r_{\rm f}$; in practice, this is very similar to its  value at the end of the evolution, i.e., $\max_r \delta S(r)\simeq \delta S(r_{\rm f})$. If $r_{\rm ud+}\lesssim 10M$, up-down binaries are still stable at the end of our evolutions and thus $\max_r \delta S\simeq 0$. If the instability onset occurs earlier, binaries start transitioning toward larger values of $\delta S$. We find that the vast majority of sources with $r_{\rm ud+}\gtrsim 50M$ are able to reach the predicted endpoint ($\max_r \delta S\gtrsim 0.95$) before the PN breakdown. As long as the instability has enough time to develop, the formal $r/M\to 0$ limit appears to provide a faithful description of dynamics. In the intermediate cases with $10M\lesssim  r_{\rm ud+}\lesssim 50M$, the instability takes places shortly before the PN breakdown and, consequently, $\delta S$ does have enough time to reach unity.  

The transition between the two regimes appears to be rather sharp, taking place over a short interval in $r$. 
To better quantify this observation, we define the instability growth ``time'' as the difference between the instability onset  $r_{\rm ud +}$ and the separation where $\delta S=0.5$, i.e.,
\begin{equation}
\delta r \equiv r_{\rm ud +}- r_{\delta S=0.5}.
\end{equation}
The bottom panels of Fig.~\ref{fig deltaS}  illustrates the behavior of $\delta r$ for the  same population of BHs. The quantity $\delta r$ can only be computed for binaries that reach $\delta S=0.5$ before the end of the evolution, thus setting the constrain $\delta r \leq r_{\rm ud +} - r_{\rm f}$. The fraction of unstable binaries (those that reach $\delta S \geq 0.5$) in this population is $35\%$. We find that the typical transition intervals are $\delta r\lesssim 100 M$, with a peak at $\delta r \simeq 25 M$, so the instability develops over a short period and unstable binaries quickly reach values of $S$ close to the endpoint.

\subsection{A simple astrophysical population} \label{A simple astrophysical population}

We now study the effect of the instability on an astrophysically-motivated population of binary BHs.
We model a formation channel that leads to the alignment of the BH spins with the orbital angular momentum, but where co-alignment and counteralignment are equally probable.
This might be the case, for instance, for stellar-mass BHs brought together by viscous interactions in AGN disks \cite{2017MNRAS.464..946S, 2017ApJ...835..165B,2018MNRAS.474.5672L, 2018ApJ...866...66M, 2019ApJ...878...85S, 2019ApJ...876..122Y, 2019PhRvL.123r1101Y, 2020MNRAS.494.1203M}.
Unlike BH binaries formed from binary stars (where the initial cloud imparts its angular momentum to both objects favoring coalignment), or systems formed in highly interacting environments like globular clusters (where frequent interactions tend to randomize the spin directions), an accretion disk defines an axisymmetric environment without a preference for co- or counteralignment.
\citeauthor{2020MNRAS.494.1203M} \cite{2020MNRAS.494.1203M} specifically modeled this scenario by assuming that 1/4 of the population is found in either the up-up, down-down, down-up, and up-down configuration.
Naively, one could expect that  $\ssim 25\%$ of the stellar-mass BH binaries formed in AGN disks are subject to the up-down instability.

As before, we distribute mass ratios using the astrophysical population inferred from the O1+O2 GW events~\cite{2019ApJ...882L..24A}, $p(q) \propto q^{6.7}$ with $q \in [0.1, 1]$, and sample the dimensionless spins $\chi_i$ uniformly in $[0.1, 1]$. We simulate $10^4$ binaries in each of the four aligned configurations, and integrate the precession equations numerically from an initial orbital separation $r_{\rm i} = 1000M$ to a final separation $r_{\rm f} = 10M$. Binaries are initialized by sampling $\cos\theta_i$ from truncated Gaussians with $\delta\theta = 20\degree$. If the corresponding parameters $q$, $\chi_1$ and $\chi_2$ are such that $r_{\mathrm{ud}+} > r_{\rm i}$ (i.e., if the source went unstable before the beginning of our integrations), the initial configuration is set to be that of a $\Delta\Phi = 0$ resonance, again with a $\delta\theta = 20\degree$ perturbation. 

\begin{figure} 
\centering
\includegraphics[width=0.85\columnwidth]{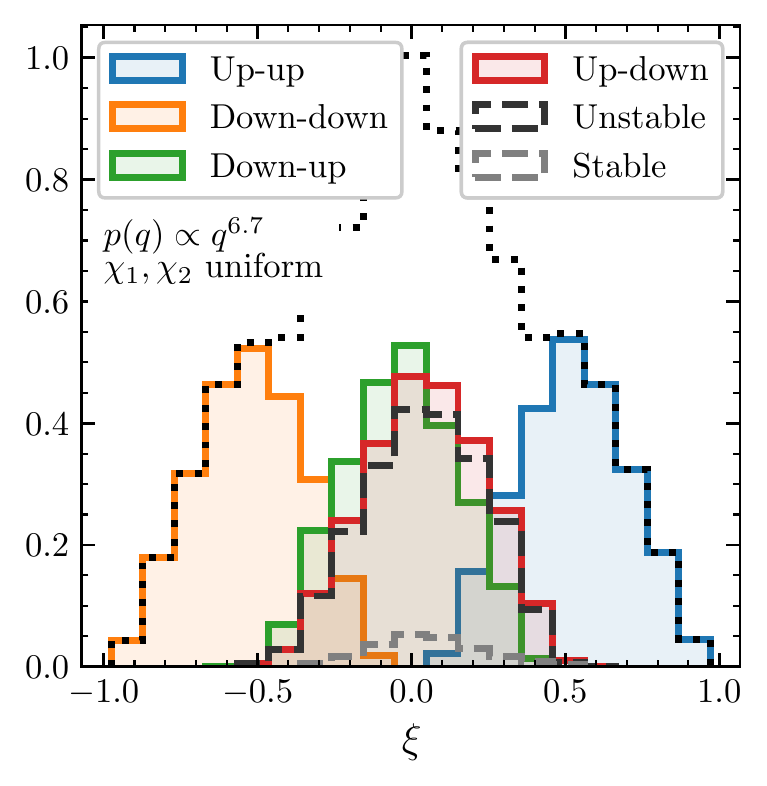}
\caption{Distribution of effective spin $\xi$ (often referred to as $\chi_{\rm eff}$) for the four populations of aligned binaries. Mass ratios $q$ are sampled according to the power law distribution $p(q) \propto q^{6.7}$ \cite{2019ApJ...882L..24A} and dimensionless spins $\chi_i$ are sampled uniformly. We set $q, \chi_1, \chi_2 \in [0.1, 1]$ and introduce an initial misalignment $\delta\theta=20^\circ$. The dotted empty histogram indicates the full population; color shaded histograms differentiate between the four aligned cases. For the up-down population (red), we further separate the contributions of two subpopulations: sources that remain stable during the entire PN inspiral ($\max_r\delta S<0.5$, dashed grey) and sources that undergo the instability  ($\max_r\delta S>0.5$, dashed black).}
\label{fig xi_hist}
\end{figure}

The resulting distribution of $\xi$ is shown in Fig.~\ref{fig xi_hist}. The effective spin $\xi$ is a constant of motion; these curves are independent of the orbital separation. Up-up (down-down) binaries tend to pile up at positive (negative) large values of the effective spins, while both up-down and down-up sources contribute to a peak at $\xi\sim 0$.

\begin{figure*}[t] 
\centering
\includegraphics[width=1.0\textwidth]{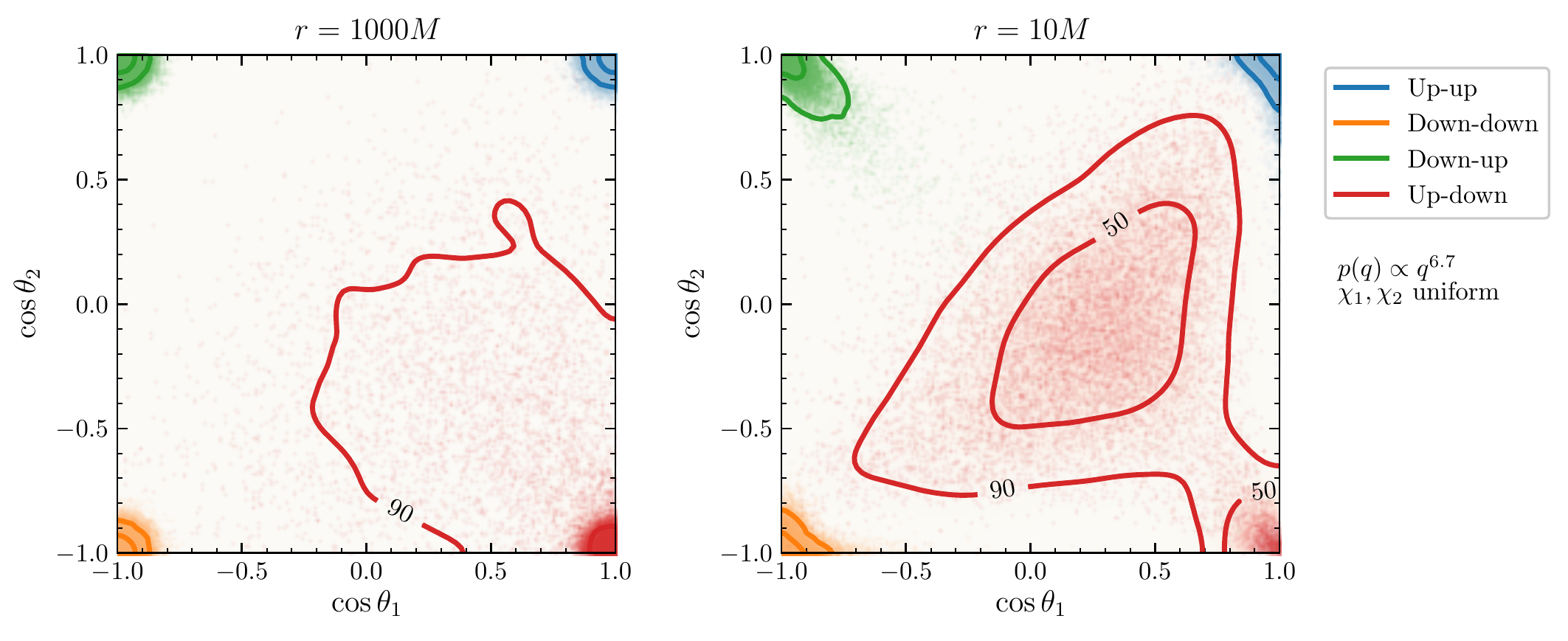}
\caption{Joint distribution of $\cos\theta_1$ and $\cos\theta_2$ for binary BHs with initially aligned spins. Sources are evolved numerically from a separation $r=1000M$ (left panel) to $r=10M$ (right panel). Mass ratios $q$ are sampled according to the power law distribution $p(q) \propto q^{6.7}$ \cite{2019ApJ...882L..24A}; dimensionless spins $\chi_i$ are sampled uniformly. We set $q, \chi_1, \chi_2 \in [0.1, 1]$.  The populations, each containing $10^4$ binaries, of up-up (blue), down-down (orange) and down-up (green) binaries remain in their initial distributions whereas the up-down (red) population does not, thus highlighting the precessional instability. By the end of the evolutions the up-down binaries split into two sub-populations: those which remain stable (bottom right corner) and those which do not (central region). The trend observed in the unstable subpopulation matches the prediction $\cos\theta_1 = \cos\theta_2$ of Sec.~\ref{Instability limit}. An animated version of this figure is available at \href{https://davidegerosa.com/spinprecession/}{www.davidegerosa.com/spinprecession}.}
\label{fig contour_cos}
\end{figure*}
Figure~\ref{fig contour_cos} shows the joint distributions of $\cos\theta_1$ and $\cos\theta_2$ at the initial (left) and final (right) separations for each of the four populations. Up-up, down-down, and down-up binaries largely retain their initial, aligned orientation. Up-down binaries segment into two clear subpopulations: those which remain stable (lower-right corner in Fig.~\ref{fig contour_cos}) and those which become unstable (center of Fig.~\ref{fig contour_cos}). The dispersion of the stable up-down binaries increases compared to the initial distribution owing to a proportion of these binaries that reach the onset of the instability but do not reach the formal endpoint by the end of the evolution. 

The subpopulation that becomes unstable presents a clear trend in the misalignment distribution: binaries pile up along the $\cos\theta_1 =\cos\theta_2$ diagonal as predicted by our  Eq.~(\ref{endpointanalytic}). As before, we characterize the two populations using $\delta S$.
At $r=1000M$, only $\ssim34\%$ binaries are in the unstable subpopulation ($ \delta S>0.5$): for the vast majority, these are the cases with $r_{\rm ud+}>r_{\rm i}$. By the time binaries reach $r=10M$, the unstable fraction goes up to $\ssim91\%$ (cf. $35\%$ for the population with mass ratios instead distributed uniformly in $[0.1, 1]$ presented in Fig.~\ref{fig deltaS}). Compared to the distribution of analytic endpoints of Fig.~\ref{fig analytic_endpoint}, the numerical population is skewed toward the initial configuration $\cos\theta_1 = - \cos\theta_2 = 1$, again due to a proportion of binaries that do not fully reach $\delta S\sim 1$.

\begin{figure*} 
\centering
\includegraphics[width=0.9\textwidth]{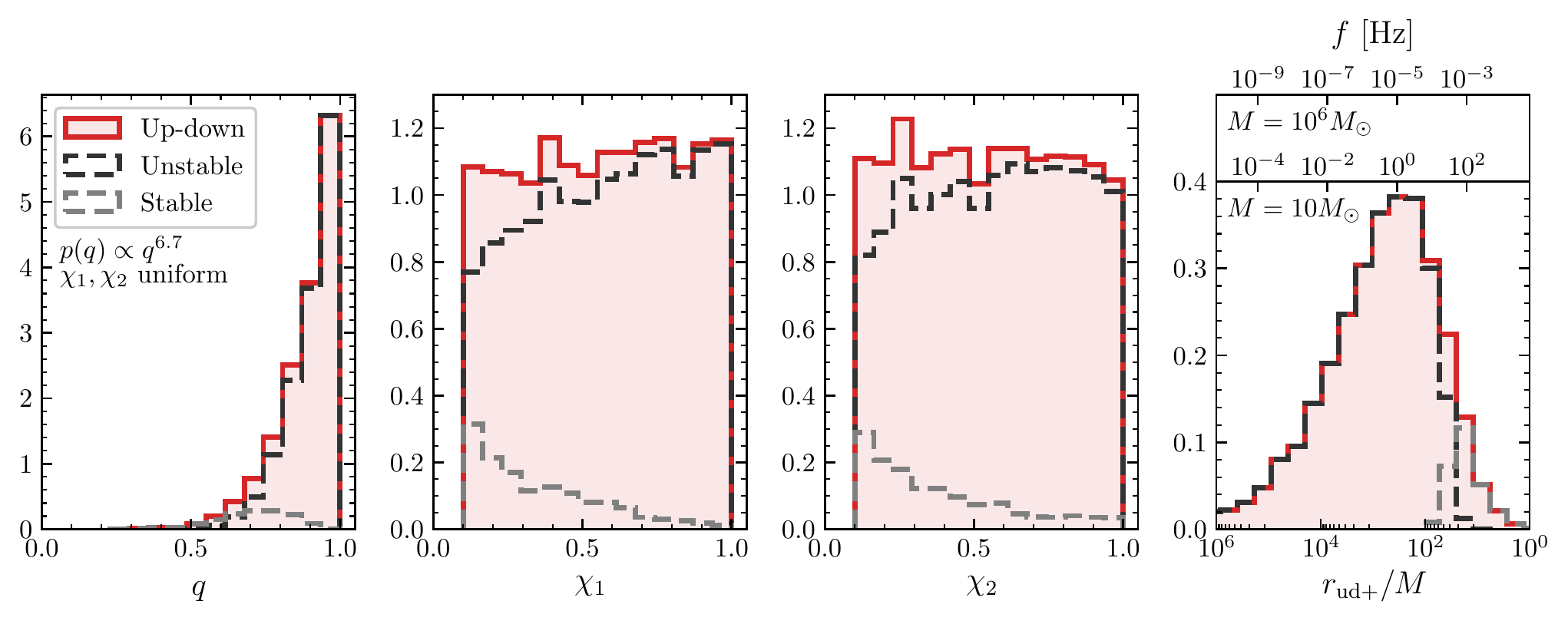}
\caption{Distribution of mass ratio $q$, spin magnitudes $\chi_i$, and instability onset $r_{\rm ud+}$ (left to right) for a set of up-down binaries. Mass ratios are sampled according to a power law distribution  that strongly favors equal masses~\cite{2019ApJ...882L..24A}, while spins are sampled uniformly. We set $q, \chi_1, \chi_2 \in [0.1, 1]$. Red histograms show the full up-down population, while dashed black (grey) histograms indicate systems that do (not) become unstable. Stability is here defined using $\max_r \delta_S \lessgtr 0.5$ and integrating from $r_{\rm i}=1000M $ to $r_{\rm f}=10M$. In the right-most panel, the top axes show the corresponding value of the GW frequency $f=\sqrt{M/\pi^2 r^3}$ for systems with total mass $M=10 M_\odot$ and $M=10^6 M_\odot$.}
\label{fig q_chi1_chi2_rudp_hists}
\end{figure*}

Figure~\ref{fig q_chi1_chi2_rudp_hists} shows the distribution of $r_{\rm ud+}$, $q$, $\chi_1$, and $\chi_2$ for the two up-down subpopulations.
Only binaries with either $q\lesssim 0.6$ or $\chi_i \lesssim 0.2$ are still stable at the end of the evolution. These values correspond to $r_{\rm ud+}\lesssim 50M$.  All other sources belong to the unstable subpopulation and approach merger near their predicted endpoints ($\max_r \delta S\gtrsim 0.5$).
An orbital separation  of $50M$ corresponds to a GW frequency $f=\sqrt{M/\pi^2 r^3}$ of $\ssim 20$ Hz for a typical LIGO source ($M=10M_\odot$) and $\ssim 10^{-4}$ Hz for a supermassive BH binary ($M=10^6 M_\odot$) detectable by LISA.

A tantalizing possibility would be the development of the precessional instability while a binary is being observed. For LIGO, we expect that such a situation is possible for only a small number of sources. To estimate this fraction, we produce a distribution of the total mass again according to Ref.~\cite{2019ApJ...882L..24A} with the distributions of $q$, $\chi_1$, and $\chi_2$ as in Fig.~\ref{fig q_chi1_chi2_rudp_hists}. The subpopulation for which the instability develops in band is then determined by the conditions $f_\mathrm{ud+} > f_\mathrm{LIGO}$ (lower LIGO frequency cutoff), $r_\mathrm{ud+} > 10M$ (validity of the PN approximation, which also provides the upper frequency cutoff) and $\max_r \delta S > 0.5$ (appreciable development of the instability). Unsurprisingly, this fraction depends strongly on the lower frequency cutoff: for $f_\mathrm{LIGO} = 20\,\mathrm{Hz} \ (10\,\mathrm{Hz})$, only $0.7\% \ (2.8\%)$ of the total population develops the precessional instability while in the LIGO band. LISA might provide better prospects, as some supermassive BH binaries will remain visible for several precession cycles~\cite{2016PhRvD..93b4003K}.

The condition $q\to 0$ and $\chi_i\to 0$ identifies the single-spin limit. In practice, we expect that the vast majority of up-down sources where two-spin effects are prominent will become unstable before entering the sensitivity window of our detectors. 
Proper modeling of two-spin effects appears to be crucial. The $\xi$ distributions of the two subpopulations does not present evident systematic trends (Fig.~\ref{fig xi_hist}) and largely reflects that of the full up-down sample. This suggests that it will be challenging to distinguish stable and unstable binaries by measuring only one effective spin.

\section{Conclusions} \label{Conclusions}

In this paper we reinvestigated the precessional instability in BH binaries first reported by \citeauthor{2015PhRvL.115n1102G}~\cite{2015PhRvL.115n1102G}. For unequal mass systems, there are four distinct configurations in which the BH spins are aligned or antialigned with the orbital angular momentum. They are all equilibrium solutions of the spin precession equations. By perturbing these configurations we tested their stability properties. While the up-up, down-down, and down-up configurations respond with stable oscillations, up-down binaries encounter an instability at orbital separations between $r_{\mathrm{ud}\pm}$, in precise agreement with the results of Ref.~\cite{2015PhRvL.115n1102G}. The instability induces precessional motion by which the two BH spins become largely misaligned with the orbital angular momentum.

We verified the occurrence of the up-down instability with numerical PN evolutions. By varying the initial misalignment of the BH spins, we found that after evolving through the instability binaries tend to cluster at a well-defined endpoint configuration, rather than dispersing in the available parameter space as might usually be expected of an instability.

The evolution toward this endpoint can be characterized in terms of the so-called spin-orbit resonances~\cite{2004PhRvD..70l4020S}. Within the framework of 2PN spin precession, we developed a semianalytic scheme to locate and identify these resonances, and proved that a binary initially in such a configuration remains so. We derived analytic solutions in the zero-separation limit and identified the asymptotic configuration of both resonant families, $\Delta\Phi=0$ and $\Delta\Phi=\pi$.

In particular, for separations $r > r_{\mathrm{ud}+}$ the up-down configuration is a $\Delta\Phi = 0$ resonance, but between $r_{\mathrm{ud}+} > r > r_{\mathrm{ud}-}$ this is no longer the case. This is precisely the cause of the precessional instability: a binary initially configured arbitrarily close to up-down is also arbitrarily close to a resonance and thus tends to remain resonant. Upon reaching the instability onset at $r_{\mathrm{ud}+}$, when up-down is no longer a resonant solution, the binary moves away from this initial alignment via precession to the new $\Delta\Phi = 0$ resonance. The asymptotic PN endpoints of binaries initialized close to up-down can therefore be found analytically. Specifically, the up-down endpoint is characterized by $\hat{\vec S}_1= \hat{\vec S}_2$ and $\hat{\vec S}_1 \cdot \hat{\vec L} =  \hat{\vec S}_2 \cdot \hat{\vec L}  = (\chi_1-q\chi_2) / (\chi_1+q\chi_2)$.

In reality, up-down binaries will not become unstable (at least in the PN regime) if $r_{\mathrm{ud}+} < 10M$ and may not fully reach their endpoint configurations if $r_{\mathrm{ud}+}$ is too small. We find that the instability develops rather quickly.
The vast majority of binaries with $r_{\mathrm{ud}+} \gtrsim 50M$ fully reach the predicted endpoint by the end of the evolution. 
More specifically, the instability develops over a characteristic separation of  $\delta r \simeq 25M$.

Because the stability-to-endpoint transition is so quick, one can further approximate the occurrence of the up-down instability as a step function. Let $r_{\rm GW}$ denote the separation at which a BH binary enters the sensitivity window of a given detector ($\ssim 10$~Hz for the case of LIGO/Virgo and $\ssim 10^{-3}$~Hz for LISA). Broadly speaking, we predict that binaries formed in the up-down configuration will be observable 
\begin{itemize}
\item still in the up-down configuration if $r_{\rm ud+}\lesssim r_{\rm GW}$; 
\item with $\vec{\hat S}_1\simeq \vec{\hat S}_2$ and  $\hat{\vec S}_1 \!\cdot\! \hat{\vec L} \simeq  \hat{\vec S}_2 \!\cdot\! \hat{\vec L}$ if $r_{\rm ud+} \gtrsim r_{\rm GW}$.
\end{itemize}

Our findings are particularly relevant for BH binary formation channels where astrophysical mechanisms tend to align the spins without preference for the alignment direction. This might be the case for stellar-mass BHs embedded in AGN accretion disks \cite{2020MNRAS.494.1203M,2019PhRvL.123r1101Y}: such a population of BHs will consist of up-up, down-down, down-up and up-down binaries in equal proportion. Over their inspirals, the distributions of the spin directions for the former three configurations remains the same. Up-down binaries, on the other hand, split into two sub-populations: those that remain stable and those that become unstable. The latter approach merger with spins coaligned with each other and equally misaligned with the orbital angular momentum, as predicted by our analytic calculation.

The analysis presented in this paper is limited to the PN regime of BH binary inspirals ($r\gtrsim 10M$). Numerical relativity simulations are necessary to fully test the instability endpoint closer to merger. Injections of up-down binaries in GW parameter-estimation tools will allow us to forecast the distinguishability of these sources with current and future interferometers. We foresee that the inclusion of two-spin effects in waveform templates will be crucial to properly characterize up-down sources. We hope that our PN predictions will spark future work in both these directions.

\acknowledgements
We thank Antoine Klein, Michael Kesden, Richard O'Shaughnessy, Emanuele Berti, Ulrich Sperhake, Christopher Moore, and Eliot Finch for discussions.
D.G. is supported by Leverhulme Trust Grant No. RPG-2019-350. Computational work was performed on the University of Birmingham BlueBEAR cluster, the Athena cluster at HPC Midlands+ funded by EPSRC Grant No. EP/P020232/1 and the Maryland Advanced Research Computing Center (MARCC).

\appendix

\section{NEAR-ALIGNMENT EXPANSION}
\label{Near-alignment expansion}

In this Appendix we derive again the threshold of the precessional instability in the small misalignment expansion using the formalism of Ref.~\cite{2013PhRvD..88l4015K}.

Given an arbitrary vector $\vec{v}$ and a direction $\hat{\vec{z}}$, one can decompose $\vec{v}$ into a component $\vec{v}_\parallel = (\vec{v} \cdot \hat{\vec{z}}) \hat{\vec{z}}$ parallel to $\hat{\vec{z}}$ and a component $\vec{v}_\perp = \vec{v} - (\vec{v} \cdot \hat{\vec{z}}) \hat{\vec{z}}$ perpendicular to $\hat{\vec{z}}$, so that $\vec{v} = \vec{v}_\parallel + \vec{v}_\perp$. If $\vec{v}$ is nearly aligned with $\hat{\vec{z}}$ then the angle $\varepsilon$ between them is small and $\vec{v} \cdot \hat{\vec{z}} = v \cos{\varepsilon} = v + \mathcal{O}(\varepsilon^2)$. Similarly, if $\vec{v}$ is nearly counteraligned with $\hat{\vec{z}}$ then we may use the small parameter $\varepsilon$ to write $\vec{v} \cdot \hat{\vec{z}} = v \cos{(\pi+\varepsilon)} = - v \cos{\varepsilon} = - v + \mathcal{O}(\varepsilon^2)$. Defining the parameter $\alpha = \pm 1$ to distinguish between coalignment ($+1$) and counteralignment ($-1$) of the vector $\vec{v}$ with respect to $\hat{\vec{z}}$, we have in either case that $\vec{v}_\parallel \approx \alpha v \hat{\vec{z}}$. The perpendicular component satisfies $|\vec{v}_\perp| = v |\sin{\varepsilon}| = v |\varepsilon| + \mathcal{O}(\varepsilon^3)$, and hence $|\vec{v}_\perp| \ll |\vec{v}_\parallel|$.

We apply this procedure to the spins $\vec{S}_1 = \vec{S}_{1\parallel} + \vec{S}_{1\perp}$ and $\vec{S}_2 = \vec{S}_{2\parallel} + \vec{S}_{2\perp}$ of the two BHs and the orbital angular momentum $\vec{L} = \vec{L}_\parallel + \vec{L}_\perp$ of the binary, using $\alpha_1$, $\alpha_2$ and $\alpha_L$ to distinguish between coalignment and counteralignment.
We neglect radiation reaction and rewrite the 2PN orbit-averaged  equations (\ref{dS1/dt}-\ref{dL/dt})  to leading order in~$\varepsilon$:
\begin{subequations}
\label{precession equation leading order}
\begin{align}
\label{dS1/dt leading order}
\frac {d \vec{S}_1} {dt} ={}& \frac {1} {2r^3} (\beta_1 \alpha_1 S_1 \vec{L}_\perp - \beta_1 \alpha_L L \vec{S}_{1\perp} + \alpha_1 S_1 \vec{S}_{2\perp} \notag \\
&\! -\alpha_2 S_2 \vec{S}_{1\perp}) \times \hat{\vec{z}} \, , \\
\label{dS2/dt leading order}
\frac {d \vec{S}_2} {dt} ={}& \frac {1} {2r^3} (\beta_2 \alpha_2 S_2 \vec{L}_\perp - \beta_2 \alpha_L L \vec{S}_{2\perp} + \alpha_2 S_2 \vec{S}_{1\perp} \notag \\
&\! -\alpha_1 S_1 \vec{S}_{2\perp}) \times \hat{\vec{z}} \, , \\
\label{dL/dt leading order}
\frac {d \vec{L}} {dt} ={}& \frac {1} {2r^3} (\beta_1 \alpha_L L \vec{S}_{1\perp} - \beta_1 \alpha_1 S_1 \vec{L}_\perp + \beta_2 \alpha_L L \vec{S}_{2\perp} \notag \\
&\! -\beta_2 \alpha_2 S_2 \vec{L}_\perp) \times \hat{\vec{z}} \, ,
\end{align}
\end{subequations}
where
\begin{subequations} \label{beta}
\begin{align}
\label{beta1}
\beta_1 &=  4 + 3q - \frac {3\alpha_L} {L} \left(q \alpha_1 S_1 +\alpha_2 S_2\right) \, , \\
\label{beta2}
\beta_2 &=  4 + \frac{3}{q} - \frac {3\alpha_L} {L} \left( \alpha_1 S_1 +\frac{\alpha_2 S_2}{q}\right) \, .
\end{align}
\end{subequations}
Completing the Cartesian frame with two additional basis vectors $\hat{\vec{x}}$ and $\hat{\vec{y}}$, one can write $\vec{L}_\perp = L_x \hat{\vec{x}} + L_y \hat{\vec{y}}$, where $L_x = \vec{L}_\perp \cdot \hat{\vec{x}}$ and $L_y = \vec{L}_\perp \cdot \hat{\vec{y}}$, and similarly for $\vec{S}_{1\perp}$ and $\vec{S}_{2\perp}$. Defining the vectors $\vec{v}_x = (S_{1x}, S_{2x}, L_x)$, $\vec{v}_y = (S_{1y}, S_{2y}, L_y)$ and $\vec{v}_z = (S_{1z}, S_{2z}, L_z)$, Eqs.~(\ref{dS1/dt leading order}-\ref{dL/dt leading order}) can now be written as
\begin{align}
\label{dv/dt}
\frac {d \vec{v}_x} {dt} = - W \vec{v}_y
\, , \quad
\frac{d \vec{v}_y} {dt} = W \vec{v}_x
\, , \quad
\frac{d \vec{v}_z} {dt} = 0
\, ,
\end{align}
where $W$ is the matrix
\begin{widetext}
\begin{align}
W = \frac {1} {2r^3}
\begin{pmatrix}
\beta_1 \alpha_L L + \alpha_2 S_2
&
- \alpha_1 S_1
&
- \beta_1 \alpha_1 S_1
\\
- \alpha_2 S_2
&
\beta_2 \alpha_L L + \alpha_1 S_1
&
- \beta_2 \alpha_2 S_2
\\
- \beta_1 \alpha_L L
&
- \beta_2 \alpha_L L
&
\beta_1 \alpha_1 S_1 + \beta_2 \alpha_2 S_2
\end{pmatrix}
\, .
\label{Wmatrix}
\end{align}
\end{widetext}
Given the conservation of $\xi$, $J$, $S_1$ and $S_2$ over the PN evolution of the binary and since we are neglecting radiation reaction ($L$ is conserved), then $dW/dt = 0$ and we can decouple the equations for $\vec{v}_x$ and $\vec{v}_y$ by taking another time derivative. This results in the following harmonic oscillator equations:
\begin{align}
\frac {d^2 \vec{v}_x} {dt^2} + W^2 \vec{v}_x = 0
\, , \quad
\frac {d^2 \vec{v}_y} {dt^2} + W^2 \vec{v}_y = 0
\, .
\end{align}
The oscillation frequencies are given by the eigenvalues of the matrix $W^2$, which are equal to the square of the eigenvalues of $W$. From Eq.~(\ref{Wmatrix}), the latter are given by $w_0=0$ and
\begin{align}
\label{eigenvalues}
w_\pm ={}& \frac{1}{4 r^3} (1+\beta_1)\alpha_1S_1 + (1+\beta_2)\alpha_2S_2 + (\beta_1+\beta_2)\alpha_LL \notag \\
&\! \pm \bigg\{ \left[ (1+\beta_1)\alpha_1S_1 + (1+\beta_2)\alpha_2S_2 + (\beta_1+\beta_2)\alpha_LL \right]^2 \notag \\
&\! -4(\beta_1\alpha_1S_1 + \beta_2\alpha_2S_2 + \beta_1\beta_2\alpha_LL) \notag \\
&\! \times (\alpha_1S_1 + \alpha_2S_2 + \alpha_LL) \bigg\}^{1/2} \, .
\end{align}
When the $w_\pm$ are real (complex), the configuration described by the parameters $\alpha_1$, $\alpha_2$ and $\alpha_L$ is stable (unstable) to precession. This behavior is determined by the argument of the square root in Eq.~(\ref{eigenvalues}).
We therefore seek the roots of this discriminant. 
Using the equalities $\alpha_1^2 = \alpha_2^2 = \alpha_L^2 = 1$ and substituting the expressions for $\beta_i$, we find that the roots of the discriminant are
\begin{subequations}
\begin{align}
L_0 &= \alpha_L \frac{q\alpha_1S_1+\alpha_2S_2}{1+q}
\, , \\
L_\pm &= \frac{q\alpha_L\alpha_1S_1 - \alpha_L\alpha_2S_2 \pm 2 \sqrt{-q\alpha_1\alpha_2S_1S_2}}{1-q}
\, .
\end{align}
\end{subequations}
The root $L_0$ is the repeated root which we have already identified as unphysical due to a corresponding value of the binary separation $r_0 \leq M$ [cf. Eq.~(\ref{L0repeated})].
As expected, in $L_\pm$ only the relative orientations of $\vec{S}_{1\parallel}$, $\vec{S}_{2\parallel}$ and $\vec{L}_\parallel$ matter and consequently the parameters $\alpha_1$, $\alpha_2$ and $\alpha_L$ appear in pairs.
As in Sec.~\ref{Binary BH spins as harmonic oscillators}, to ensure that $L_\pm$ is real and non-negative we require $\alpha_1 \alpha_L =-\alpha_2 \alpha_L= -\alpha_1 \alpha_2 =1$, which corresponds to the up-down configuration.
We thus recover the binary separations that determine the threshold of the up-down instability, cf. Eq.~(\ref{rud}) and Ref.~\cite{2015PhRvL.115n1102G}:
\begin{align}
r_{\mathrm{ud}\pm} = \frac{\left( \sqrt{\chi_1} \pm \sqrt{q\chi_2} \right)^4}{(1-q)^2} M \, .
\end{align}

\section{COEFFICIENTS OF $\Sigma(S^2)$ AND $\Delta(J^2)$}
\label{longappendix}

\begingroup
\allowdisplaybreaks[4]

For completeness, in this Appendix we report  some of the expressions that were omitted from the main body of the paper.

The coefficients $\sigma_i$ of the third-degree polynomial $\Sigma(S^2)$ given in Eq.~(\ref{S2 cubic}) are
\begin{subequations}
\begin{align}
\sigma_6 ={}& q (1+q)^2 \, , \\ 	
\sigma_4 ={}& (1+q)^2 \big[-2 J^2 q+L^2 \big(1+q^2\big) \notag \\
&\! +2 L M^2 \xi q -(1-q) \big(qS_1^2-S_2^2\big)\big] \, , \\
\sigma_2 ={}&	2\big(1+q\big)^2 \big(1-q\big) \big[	J^2 \big(qS_1^2-S_2^2\big) \notag \\
&\! -L^2 \big(S_1^2-qS_2^2\big) \big] + q \big(1+q\big)^2 \big(J^2-L^2\big)^2 \notag \\
&\! -2LM^2\xi q \big(1+q\big) \big[\big(1+q\big) \big(J^2-L^2\big) \notag \\
&\!+\big(1-q\big) \big(S_1^2-S_2^2\big) \big] + 4L^2M^4\xi^2 q^2 \, , \\
\sigma_0 ={}&	\big(1-q^2\big) \big[	L^2 \big(1-q^2\big) \big(S_1^2-S_2^2\big)^2 \notag \\
&\! -\big(1+q\big) \big(qS_1^2-S_2\big)^2 \big(J^2-L^2\big)^2 \notag \\
&\! +2LM^2\xi q \big(S_1^2-S_2^2\big) \big(J^2-L^2\big)\big] \, .
\end{align}
\end{subequations}
These expressions were also reported in Eq.~(16) of Ref.~\cite{2015PhRvD..92f4016G}. 

The coefficients $\delta_i$ of the discriminant $\Delta(J^2)$ given in Eq.~(\ref{cubic discriminant J2}) are

\begin{widetext}
\begin{subequations}
\begin{align}
\delta_{10} ={}&\! -4 L^2 (q-1)^2 q^3 (q+1)^8
\, , \\
\delta_8 ={}& L^2 (q-1)^2 q^2 (q+1)^6 \big[L^2 \big(q^2+18 q+1\big) (q+1)^2+20 L M^2 \xi  q (q+1)^2+4q^2 \big(M^4 \xi ^2+7 S_1^2+7 S_2^2\big)-12 q^4 S_1^2
\notag \\
&\! -12 S_2^2 -4q^3 \big(S_1^2-5 S_2^2\big)+ 4q \big(5 S_1^2-S_2^2\big)\big]
\, , \\
\delta_6 ={}&\! -4 L^2 (q-1)^2 q (q+1)^6 \big\{L^4 (q+1)^2 \big(q^2+8 q+1\big) q+L^3 M^2 \xi  (q+1)^2 \big(q^2+18 q+1\big) q+L^2 \big[q^4 \big(8 M^4 \xi ^2+6 S_1^2
\notag \\
&\! +15 S_2^2\big) +q^3 \big(28 M^4 \xi^2+26 S_1^2+26 S_2^2\big)+q^2 \big(8 M^4 \xi ^2+15 S_1^2+6 S_2^2\big)-5 q^6 S_1^2+q^5 \big(S_2^2-11 S_1^2\big)+q \big(S_1^2-11 S_2^2\big)
\notag \\
&\! -5S_2^2\big] +L M^2 \xi  q \big[q^2 \big(4 M^4 \xi ^2+17 S_1^2+17 S_2^2\big)-q^4 S_1^2 -4q^3 \big(S_1^2-5 S_2^2\big)+4 q \big(5S_1^2-S_2^2\big)-S_2^2\big]
\notag \\
&\! -q^4 \big[2 S_1^2 \big(M^4 \xi^2+S_2^2\big) +11 S_1^4-10 S_2^4\big]+4 q^3 \big[S_1^2 \big(M^4 \xi ^2+5S_2^2\big)+M^4 \xi ^2 S_2^2+2 S_1^4+2 S_2^4\big]+q^2 \big(-2 M^4 \xi ^2 S_2^2
\notag \\
&\! +10 S_1^4-2 S_1^2 S_2^2-11 S_2^4\big) 3 q^6 S_1^4-6 q^5 \big(S_1^4+2 S_1^2 S_2^2 \big) -6q \big(2 S_1^2 S_2^2+S_2^4\big)+3 S_2^4\big\}
\, , \\
\delta_4 ={}& 2 L^2 (q+1)^2 \big(q^2-1\big)^2 \big\{q^2 (q+1)^4 \big(3 q^2+14 q+3\big) L^6+6 M^2 q^2 (q+1)^4 \big(q^2+8 q+1\big) \xi  L^5-2 (q+1)^2 \big[S_1^2 q^8
\notag \\
&\! +13 S_1^2 q^7 +\big(-\xi ^2 M^4+14S_1^2-2 S_2^2\big) q^6-\big(26 \xi ^2 M^4+20 S_1^2+17 S_2^2\big) q^5-\big(72 \xi ^2 M^4+37 S_1^2+37 S_2^2\big) q^4
\notag \\
&\! -\big(26 \xi ^2 M^4+17 S_1^2+20S_2^2\big) q^3-\big(\xi ^2 M^4+2 S_1^2-14 S_2^2\big) q^2 +13 S_2^2 q+S_2^2\big] L^4 - 2 M^2 q (q+1)^2 \xi \big[4 S_1^2 q^6+\big(22 S_1^2
\notag \\
&\! -3S_2^2\big) q^5-2 \big(2 \xi ^2 M^4+9 S_1^2+17 S_2^2\big) q^4-\big(44 \xi ^2 M^4+67 S_1^2 +67 S_2^2\big) q^3-2 \big(2 \xi ^2 M^4+17 S_1^2+9 S_2^2\big) q^2+\big(22 S_2^2
\notag \\
&\! -3 S_1^2\big) q+4 S_2^2\big] L^3 + \big\{-4 S_1^4 q^{10}-10 \big(2 S_1^4+3 S_2^2 S_1^2\big) q^9-\big[53 S_1^4+\big(4 \xi ^2 M^4+74S_2^2\big) S_1^2-3 S_2^4\big] q^8+4 \big[12 S_2^2 \xi ^2 M^4
\notag \\
&\! -17 S_1^4+9 S_2^4+S_1^2 \big(6 M^4 \xi ^2-10 S_2^2\big)\big] q^7+\big[8 \xi ^4 M^8+128 S_2^2 \xi ^2 M^4+S_1^4+101 S_2^4+S_1^2 \big(92 \xi ^2 M^4+26 S_2^2\big)\big] q^6
\notag \\
&\! +4 \big[8 \xi ^4 M^8 +36 S_2^2 \xi ^2 M^4+25 S_1^4+25S_2^4+S_1^2 \big(36 \xi ^2 M^4+11 S_2^2\big)\big] q^5+\big[8 \xi ^4 M^8+92 S_2^2 \xi ^2 M^4+101 S_1^4+S_2^4
\notag \\
&\! +2 S_1^2 \big(64 \xi ^2 M^4+13S_2^2\big)\big] q^4+4 \big[6 S_2^2 \xi ^2 M^4+9 S_1^4-17 S_2^4-2 S_1^2 \big(5 S_2^2-6 	M^4 \xi ^2\big)\big] q^3+\big(-4 S_2^2 \xi ^2 M^4+3S_1^4 -53 S_2^4
\notag \\
&\! -74 S_1^2 S_2^2\big) q^2 -10 \big(2 S_2^4+3 S_1^2 S_2^2\big) q-4 S_2^4\big\} L^2-2 M^2 q (q+1) \xi  \big\{4 S_1^4 q^7+	\big(3S_1^2 S_2^2-11 S_1^4\big) q^6+\big[5 S_1^4+\big(9 S_2^2
\notag \\
&\! -4 M^4 \xi ^2\big) S_1^2 -30 S_2^4\big] q^5+\big[29 S_1^4+4 \big(\xi ^2 M^4+3	S_2^2\big) S_1^2-3 \big(4 S_2^2 \xi ^2 M^4+7 S_2^4\big)\big] q^4+\big[4 S_2^2 \xi ^2 M^4-21S_1^4+29 S_2^4
\notag \\
&\! +12 S_1^2 \big(S_2^2 -M^4 \xi^2\big)\big] q^3+\big(-4 S_2^2 \xi ^2 M^4-30 S_1^4+5 S_2^4+9 S_1^2 S_2^2\big) q^2+\big(3 S_1^2 S_2^2-11 S_2^4\big) q+4 S_2^4\big\} L-2 (q+1)^2
\notag \\
&\! \times \big\{S_1^6 q^8-\big(7 S_1^6 +9 S_2^2 S_1^4\big) q^7+\big[S_1^6+\big(9 S_2^2-M^4 \xi ^2\big) S_1^4+18 S_2^4S_1^2\big] 	q^6+\big[17S_1^6+\big(6 \xi ^2 M^4+9 S_2^2\big)S_1^4 -10 S_2^6
\notag \\
&\! +6 M^4 S_2^2 \xi ^2 S_1^2\big] q^5 -\big[2 S_1^6+3 \big(2 \xi ^2 M^4+9 S_2^2\big) S_1^4+\big(10 S_2^2 \xi ^2 M^4+27 S_2^4\big) 	S_1^2+2 S_2^4 \big(3 \xi ^2 M^4+S_2^2\big)\big] q^4 +\big[-10 S_1^6
\notag \\
&\! +\big(6 S_2^2 \xi ^2 M^4+9 S_2^4\big) S_1^2+17 S_2^6+6 M^4 S_2^4 \xi ^2\big] q^3+\big(S_2^6+9 S_1^2 S_2^4-M^4 \xi ^2 S_2^4+18 S_1^4 S_2^2\big) q^2-\big(7S_2^6+9 S_1^2 S_2^4\big) q
\notag \\
&\! +S_2^6\big\}\big\}
\, , \\
\delta_2 ={}&\! -4 L^2 (q+1)^2 \big(q^2-1\big)^2 \big\{q^2 (q+1)^4 \big(q^2+3 q+1\big) L^8+M^2 q^2 (q+1)^4 \big(3q^2+14 q+3\big) \xi  L^7-(q+1)^2 \big[2 S_1^2 q^8
\notag \\
&\! +11 S_1^2 q^7+\big(-2 \xi ^2 M^4+7 S_1^2-S_2^2\big) q^6-\big(28 \xi ^2 M^4+18 S_1^2+ 9S_2^2\big) q^5-8 \big(8 \xi ^2 M^4+3 S_1^2+3 S_2^2\big) q^4-\big(28 \xi ^2 M^4
\notag \\
&\! +9 S_1^2+18S_2^2\big) q^3-\big(2 \xi ^2 M^4+S_1^2-7 S_2^2\big) q^2+11 S_2^2 q + 2S_2^2\big] L^6-M^2 q (q+1)^2 \xi  \big[12 S_1^2 q^6+\big(29 S_1^2-2S_2^2\big) q^5
\notag \\
&\! -4 \big(3 \xi ^2 M^4+10 S_1^2+5 S_2^2\big) q^4-\big(68 \xi ^2 M^4+75 S_1^2+75 S_2^2\big) q^3-4 \big(3 \xi ^2 M^4+5 S_1^2+10 S_2^2\big) q^2+\big(29 S_2^2-2 S_1^2\big) q
\notag \\
&\! +12 S_2^2\big] L^5+\big\{-2 S_1^2 \big(3 S_1^2+S_2^2\big) q^{10}-\big(29 S_1^4+6 S_2^2 S_1^2\big) q^9+\big[2 S_2^2 \xi ^2 M^4-45 S_1^4+S_2^4-2 S_1^2 \big(9 \xi ^2 M^4+16 S_2^2\big)\big] q^8
\notag \\
&\! -2 \big[4 S_1^4+\big(3 \xi ^2 M^4+29 S_2^2\big) S_1^2-3 \big(2 S_2^2 \xi ^2 M^4+S_2^4\big)\big] q^7+\big[24 \xi ^4 M^8+98 S_2^2 \xi ^2 M^4+49 S_1^4+25 S_2^4+2 S_1^2 \big(61 \xi ^2 M^4
\notag \\
&\! +5 S_2^2\big)\big] q^6+\big[80 \xi ^4 M^8+198 S_2^2 \xi ^2 M^4+55 S_1^4+55 S_2^4+2 S_1^2 	\big(99 \xi ^2 M^4+40 S_2^2\big)\big] q^5+\big[24 \xi ^4 M^8+122 S_2^2 \xi ^2 M^4
\notag \\
&\! +25 S_1^4+49 S_2^4+2 S_1^2 \big(49 \xi ^2 M^4+5 S_2^2\big)\big] q^4+2 \big[-3 S_2^2 \xi ^2 M^4+3 S_1^4-4 S_2^4+S_1^2 \big(6 M^4 \xi ^2-29 S_2^2\big)\big] q^3+\big[S_1^4
\notag \\
&\! +\big(2 M^4 \xi ^2-32 S_2^2\big) S_1^2-9 \big(2 S_2^2 \xi ^2 M^4+5 S_2^4\big)\big] q^2-\big(29 S_2^4+6 S_1^2 S_2^2\big) q-2 S_2^2 \big(S_1^2+3 S_2^2\big)\big\} L^4+M^2 q \xi  \big\{-8 S_1^2 \big(3 S_1^2
\notag \\
&\! +S_2^2\big) q^8+\big(-51 S_1^4-8 S_2^2 S_1^2+3 S_2^4\big) q^7+\big[8 S_2^2 \xi ^2 M^4-2 S_1^4+14 S_2^4+4 S_1^2 \big(2 \xi ^2 M^4+S_2^2\big)\big] q^6+\big[12 S_2^2 \xi ^2 M^4
\notag \\
&\! +65 S_1^4+31 S_2^4+S_1^2 \big(76 M^4 \xi ^2-40 S_2^2\big)\big] q^5+4 \big[4 \xi ^4 M^8+18 S_2^2 \xi ^2 M^4+15 S_1^4+15 S_2^4+S_1^2 \big(18 M^4 \xi ^2-22S_2^2\big)\big] q^4 \notag\\
&\! +\big[76 S_2^2 \xi ^2 M^4+31 S_1^4+65 S_2^4+S_1^2 \big(12 M^4 \xi ^2-40 S_2^2\big)\big] q^3+2 \big[4 S_2^2 \xi ^2 M^4+7S_1^4-S_2^4+2 S_1^2 \big(2 \xi ^2 M^4+S_2^2\big)\big] q^2
\notag \\
&\! +\big(3 S_1^4-8 S_2^2 S_1^2-51 S_2^4\big) q-8 S_2^2 \big(S_1^2+3 S_2^2\big)\big\} L^3-\big\{\big(6 S_1^6+4 S_2^2 S_1^4\big) q^{10}+\big(13 S_1^6+8 S_2^2 S_1^4+15 S_2^4 S_1^2\big) q^9+\big[S_1^6
\notag \\
&\! +\big(18 M^4 \xi ^2-41 S_2^2\big) S_1^4+\big(4 S_2^2 \xi ^2 M^4+11 S_2^4\big) S_1^2-S_2^6\big] q^8-\big[9 S_1^6+3 \big(8 \xi ^2 M^4+31 S_2^2\big) S_1^4+\big(12 S_2^2 \xi ^2 M^4 +35 S_2^4\big)
\notag \\
&\! \times S_1^2 +7 S_2^6+24 M^4 S_2^4 \xi ^2\big] q^7+\big[-8 S_2^2 \xi ^4 M^8+4 S_2^4 \xi^2 M^4+3 S_1^6-9 S_2^6+S_1^4 \big(9 S_2^2-50 M^4 \xi ^2\big)+S_1^2 \big(-24 \xi ^4 M^8
\notag \\
&\! +24 S_2^2 \xi ^2 M^4+17 S_2^4\big)\big] q^6+\big[8 S_2^2 \xi^4 M^8+20 S_2^4 \xi ^2 M^4+3 S_1^6+3 S_2^6+5 S_1^4 \big(4 \xi ^2 M^4+21 S_2^2\big)+S_1^2 \big(8 \xi ^4 M^8
\notag \\
&\! +80 S_2^2 \xi ^2 M^4+105 S_2^4\big)\big] q^5+\big[-24 S_2^2 \xi ^4 M^8-50 S_2^4 \xi ^2 M^4-9 S_1^6+3 S_2^6+S_1^4 \big(4 \xi ^2 M^4+17 S_2^2\big)+S_1^2 \big(-8 \xi ^4 M^8
\notag \\
&\! +24 S_2^2 \xi ^2 M^4+9 S_2^4\big)\big] q^4-\big[7 S_1^6+\big(24 \xi ^2 M^4+35 S_2^2\big) S_1^4+3 \big(4 S_2^2 \xi ^2 M^4+31 S_2^4\big) S_1^2+9 S_2^6+24 M^4 S_2^4 \xi ^2\big] q^3
\notag \\
&\! +\big[-S_1^6+11 S_2^2 S_1^4+\big(4 M^4 S_2^2 \xi ^2-41 S_2^4\big) S_1^2+S_2^6+18 M^4 S_2^4 \xi ^2\big] q^2+\big(13 S_2^6+8S_1^2 S_2^4+15 S_1^4 S_2^2\big) q+6 S_2^6
\notag \\
&\! +4 S_1^2 S_2^4\big\} L^2-M^2 q (q+1) \xi  \big\{4 \big(3 S_1^6+2 S_2^2 S_1^4\big) q^7+\big(-17 S_1^6-6 S_2^2 S_1^4+3 S_2^4 S_1^2\big) q^6-\big[21 S_1^6+2 \big(6 \xi ^2 M^4+11 S_2^2\big)
\notag \\
&\! \times S_1^4 +\big(8 M^4 S_2^2 \xi ^2-3 S_2^4\big)   S_1^2+20 S_2^6\big] q^5+\big[37 S_1^6+\big(20 \xi ^2 M^4+3 S_2^2\big) S_1^4+\big(12 S_2^2 \xi ^2 M^4+11 S_2^4\big) S_1^2+9 S_2^6
\notag \\
&\! -12 M^4 S_2^4 \xi ^2\big] q^4+\big[9 S_1^6+\big(11 S_2^2-12 M^4 \xi ^2\big) S_1^4+3 \big(4 S_2^2 \xi ^2 M^4+S_2^4\big) S_1^2+37 S_2^6+20 M^4 S_2^4 \xi ^2\big] q^3-\big[20 S_1^6
\notag \\
&\! -3 S_2^2 S_1^4+\big(8 S_2^2 \xi ^2 M^4+22 S_2^4\big) S_1^2+3 S_2^4 \big(4 \xi ^2 M^4+7 S_2^2\big)\big] q^2+\big(-17S_2^6-6 S_1^2 S_2^4+3 S_1^4 S_2^2\big) q+12 S_2^6+8 S_1^2 S_2^4\big\} L
\notag \\
&\! -(q-1) (q+1)^2 \big(q S_1^2-S_2^2\big) \big\{2 S_1^4 \big(S_1^2+S_2^2\big) q^6+\big(-3 S_1^6+2 S_2^2 S_1^4-7 S_2^4 S_1^2\big) q^5-\big[7 S_1^6+\big(2 M^4 \xi ^2-5 S_2^2\big) S_1^4
\notag \\
&\! +\big(2S_2^2 \xi ^2 M^4+7 S_2^4\big) S_1^2-5 S_2^6\big] q^4+\big[3 S_1^6+\big(4 \xi ^2 M^4+5 S_2^2\big) S_1^4+5 S_2^4 S_1^2+3 S_2^6+4 M^4 S_2^4 \xi ^2\big] q^3+\big[5 S_1^6-7 S_2^2 S_1^4
\notag \\
&\! +\big(5 S_2^4-2 M^4 S_2^2 \xi ^2\big) S_1^2-7 S_2^6-2 M^4 S_2^4 \xi ^2\big] q^2+\big(-3S_2^6+2 S_1^2 S_2^4-7 S_1^4 S_2^2\big) q+2 S_2^4 \big(S_1^2+S_2^2\big)\big\}\big\}
\, , \\
\delta_0 ={}& L^2 (q-1)^2 (q+1)^2 \big[L^2 (q+1)^2+2 L M^2 \xi  q+\big(q^2-1\big) \big(S_1^2-S_2^2\big)\big]^2 \big\{L^6 q^2 (q+1)^4+4 L^5 M^2 \xi  q^2 (q+1)^4
\notag \\
&\! -2 L^4 (q+1)^2 \big[-q^4 \big(2 M^4 \xi^2+3 S_1^2+S_2^2\big)-4 q^3 \big(2 M^4 \xi^2 +S_1^2+S_2^2\big)-q^2 \big(2 M^4 \xi ^2+S_1^2+3 S_2^2\big)+2 q^6 S_1^2+2 q^5 S_1^2
\notag \\
&\! +2 q S_2^2+2S_2^2\big]+4 L^3 M^2 \xi  q (q+1)^2 \big[q^2 \big(4 M^4 \xi ^2+7 S_1^2+7 S_2^2\big)-4 q^4 S_1^2+q^3 \big(2 S_1^2+S_2^2\big)+q \big(S_1^2+2S_2^2\big)-4 S_2^2\big]
\notag \\
&\! +L^2 \big\{q^6 \big[-2 S_1^2 \big(4 M^4 \xi ^2+21 S_2^2\big)-15 S_1^4+S_2^4\big]+4 q^5 \big[S_1^2 \big(4 M^4 \xi ^2+3 S_2^2\big)+8 M^4 \xi ^2 S_2^2+5 S_1^4+3 S_2^4\big]
\notag \\
&\! +4 q^3 \big[S_1^2 \big(8 M^4 \xi ^2+3 S_2^2\big)+4 M^4 \xi ^2 S_2^2+3 S_1^4+5S_2^4\big]+q^2 \big(-8 M^4 \xi ^2 S_2^2+S_1^4-42 S_1^2 S_2^2-15 S_2^4\big)+2 q^4 \big[8 M^8 \xi^4 
\notag \\
&\! +S_1^2 \big(28 M^4 \xi ^2+34 S_2^2\big)+28 M^4 \xi ^2 S_2^2+15 S_1^4+15 S_2^4\big]-8q^8 S_1^4-4 q^7 \big(6 S_1^4+5 S_1^2 S_2^2\big)-4 q \big(5 S_1^2 S_2^2+6 S_2^4\big)-8S_2^4\big\}
\notag \\
&\! -4 L M^2 \xi  q (q+1) \big\{-q^3 \big[S_1^2 \big(4 M^4 \xi ^2-3 S_2^2\big)+6 S_1^4+5 S_2^4\big]-	q^2 \big(4 M^4 \xi ^2 S_2^2+5 S_1^4-3 S_1^2 S_2^2+6 S_2^4\big)+4 q^5 S_1^4
\notag \\
&\! +q^4 S_1^2 \big(3 S_1^2+S_2^2\big)+q S_2^2 \big(S_1^2+3 S_2^2\big)+4 S_2^4\big\}-4 (q+1)^2 \big(S_2^2-q S_1^2\big)^2 \big[-q^2 \big(M^4 \xi ^2+S_1^2+S_2^2\big)+q^4 S_1^2
\notag \\
&\! +q^3 \big(S_1^2-S_2^2\big)+q \big(S_2^2-S_1^2\big)+S_2^2\big]\big\}
\, . \\
\nonumber
\end{align}
\end{subequations}
\end{widetext}
\endgroup

\balancecolsandclearpage

\bibliography{updown_instability}

\end{document}